\documentclass[twocolumn,english,prb]{revtex4}
\usepackage[T1]{fontenc}
\usepackage[latin9]{inputenc}
\usepackage{float}
\usepackage{amsmath}
\usepackage{amssymb}
\usepackage{graphicx}
\usepackage{esint}

\makeatletter
\@ifundefined{textcolor}{}
{%
 \definecolor{BLACK}{gray}{0}
 \definecolor{WHITE}{gray}{1}
 \definecolor{RED}{rgb}{1,0,0}
 \definecolor{GREEN}{rgb}{0,1,0}
 \definecolor{BLUE}{rgb}{0,0,1}
 \definecolor{CYAN}{cmyk}{1,0,0,0}
 \definecolor{MAGENTA}{cmyk}{0,1,0,0}
 \definecolor{YELLOW}{cmyk}{0,0,1,0}
}

\usepackage{babel}

\makeatother

\usepackage{babel}
\begin{document}
\title{Correlation effects in spin models in the presence of a spin bath}
\author{Álvaro Gómez-León$^{1}$, Tim Cox$^{2}$ and Philip Stamp$^{2}$}
\affiliation{$^{1}$Instituto de Ciencia de Materiales, CSIC, Cantoblanco, Madrid
E-28049, Spain}
\affiliation{$^{2}$Department of Physics and Astronomy, University of British
Columbia, 6224 Agricultural Road, Vancouver, British Columbia, V6T
1Z1, Canada}
\date{\today}
\begin{abstract}
We analyze the effect of a bath of spins interacting with a spin system
in terms of the equation of motion technique. We show that this formalism
can be used with general spin systems and baths, and discuss the concrete
case of a Quantum Ising model longitudinally coupled to the bath.
We show how the uncorrelated solutions change when spin-spin correlations
are included, the properties of the quasiparticle excitations and
the effect of internal dynamics in the spin bath.
\end{abstract}
\maketitle

\section{Introduction}

Spins systems are among the most studied models in physics, but their
properties are also closely related with systems in many other fields
of science \cite{VoterModel,Biophysics,BirdNavigation}. These models
are interesting because they display many emergent phenomena due to
collective effects, and in addition, because of their relevance in
the development of quantum computers\cite{QAnnealing}. Most works
on spin models mainly consider closed systems and study the equilibrium
phase diagram, the critical exponents, and more recently, also the
quench dynamics\cite{QuenchDynamics-PhaseTrans,sachdev_2011}. Less
it is known about these models in the presence of environments, specially
for those described by a collection of localized modes. These are
known to produce radically different effects than the usual oscillator
bath in certain regimes \cite{LiHoFPhaseTransition,HyperfineLiHoF,Theory-SpinBath}.
As the coupling to localized spin environments is generally not weak\footnote{Typically oscillator bath couplings scale as $1/\sqrt{N}$ due to
their delocalized nature, being $N$ the number of modes. Therefore
perturbative calculations can, in most cases, capture the main properties.}, the calculations are generally non-perturbative and must be done
carefully, which implies that in the case of quantum models, the effect
of correlations can become quite important.

For practical purposes, the study of spin models has recently become
more important due to the development of the first quantum simulators
and quantum computers \cite{D-wave,Google-QC,IBM}. In this case,
environments are ubiquitous and its interaction with the quantum computer
during the computation time needs to be properly understood -- e.g.,
the fate of the quantum critical point (QCP) during the quantum annealing
process, in the presence of an environment. In the case of quantum
computers made of flux qubits, spin environments can be produced by
paramagnetic impurities and nuclear isotopes in the substrate, and
even if they weakly couple, their effect in the dynamics can be non-trivial
\cite{SQUID-dyn}. Also it has been shown that for superconducting
qubits dielectric loss can dominate, and it is produced by effective
two level systems \cite{MartinisDecoherenceSquid}.

To understand the physical effect of spin bath environments we study
a spin model coupled to a set of localized spins. We use the equation
of motion (EOM) technique for the Green's functions, and a decoupling
scheme based on a hierarchical expansion in terms of correlations.
This approach transforms the EOM into a Dyson's equation for the Green's
function, with explicit non-perturbative expressions for the self-energy,
which include a full frequency and momentum dependence. Hence it can
also be used to study quantum dynamics of excited states, as already
shown in \cite{AnalyticalDecoherence}. The lowest order solution
agrees with mean-field (MF) treatments and can explain some general
features of the effect of a baths of spins, as the solutions are solely
characterized by the connectivity between Ising spins and bath spins
(i.e., the three different coordination numbers that one can define
for this model). Higher order corrections depend on the specific details
of the bath, however, we numerically solve the self-consistent equations
for some specific cases and demonstrate that quantum correlations
tend to suppress the effect of the bath of spins, with respect to
the MF calculations.

\section{Model and method}

We consider a model with a central system and a bath. The central
system is made of a set of interacting spins coupled to an external
field $\vec{B}$. Similarly, the bath is made of a set of spins as
well, but we assume that these are non-interacting, although they
can be affected by an external field $\vec{\Delta}$\footnote{Interactions between bath spins can be easily introduced, but it is
known that large interactions would make the environment behave as
an effective oscillator bath, as the modes would not be localized
anymore.}. The total Hamiltonian is $H=H_{S}+H_{B}$, where $H_{S}$ and $H_{B}$
are given by (Greek indices specify the direction of the spin vector
and latin indices label the different sites):

\begin{eqnarray}
H_{S} & = & -\sum_{\mu,i}B_{\mu}S_{i}^{\mu}-\sum_{\mu}\sum_{i,j>i}V_{i,j}^{\mu}S_{i}^{\mu}S_{j}^{\mu}\\
H_{B} & = & -\sum_{\mu,l}\Delta_{\mu}I_{l}^{\mu}+\sum_{\mu}\sum_{i,l}A_{i,l}^{\mu}S_{i}^{\mu}I_{l}^{\mu}
\end{eqnarray}
$H_{S}$ corresponds to a general spin model coupled to an external
field $\vec{B}$ and with arbitrary interaction $V_{i,j}^{\mu}$,
while $H_{B}$ corresponds to the bath Hamiltonian which couples to
a field $\vec{\Delta}$ and interacts with the central system via
$A_{i,l}^{\mu}$. To study the properties of the system we define
the next double-time Green's functions \cite{Zubarev}: 
\begin{eqnarray}
G_{n,m}^{\alpha,\beta}\left(t,t^{\prime}\right) & = & -i\langle S_{n}^{\alpha}\left(t\right);O_{m}^{\beta}\left(t^{\prime}\right)\rangle\\
Y_{n,m}^{\alpha,\beta}\left(t,t^{\prime}\right) & = & -i\langle I_{n}^{\alpha}\left(t\right);O_{m}^{\beta}\left(t^{\prime}\right)\rangle
\end{eqnarray}
where $;$ indicates that the Green's functions may be a time ordered,
retarded or advanced. Also $O_{m}^{\beta}$ is some general spin operator
at site $m$ (i.e., it can represent a system or a bath spin). The
equation of motion for the Green's function is:
\begin{eqnarray}
i\partial_{t}G_{n,m}^{\alpha,\beta} & = & \delta\left(t-t^{\prime}\right)\langle\left[S_{n}^{\alpha},O_{m}^{\beta}\right]_{\pm}\rangle+i\epsilon_{\mu\alpha\rho}B_{\mu}G_{n,m}^{\rho,\beta}\label{eq:EOM-G1}\\
 &  & +i\epsilon_{\mu\alpha\rho}\left(\sum_{i\neq n}V_{i,n}^{\mu}G_{in,m}^{\mu\rho,\beta}-\sum_{l}A_{n,l}^{\mu}Y_{ln,m}^{\mu\rho,\beta}\right)\nonumber 
\end{eqnarray}
where $\left[\ldots\right]_{\pm}$ refers to anti-commutator or commutator,
respectively\cite{Fermionic-GF,SCHREIBER197927} (i.e., to a fermionic
or bosonic Green's function). We have defined the three-spin correlators
$G_{in,m}^{\mu\rho,\beta}\left(t,t^{\prime}\right)=-i\langle S_{i}^{\mu}S_{n}^{\rho};O_{m}^{\beta}\rangle$
and $Y_{ln,m}^{\nu\rho,\beta}\left(t,t^{\prime}\right)=-i\langle I_{l}^{\nu}S_{n}^{\rho};O_{m}^{\beta}\rangle$.

An important feature of many-body systems is the hierarchical structure
of the EOM, where interaction terms produce higher-order Green's functions.
To deal with this hierarchy, one needs to devise a method to decouple
the infinite set of equations into smaller blocks, which correctly
captures the regime of interest. For that purpose we separate the
Green's functions into their uncorrelated and correlated parts: $G_{in,m}^{\mu\rho,\beta}=\langle S_{i}^{\mu}\rangle G_{n,m}^{\rho,\beta}+\langle S_{n}^{\rho}\rangle G_{i,m}^{\mu,\beta}+\mathcal{G}_{in,m}^{\mu\rho,\beta}$
(we will use calligraphic letters to denote the correlated parts).
This separation is completely general and just transforms the initial
EOM for the two-point function in two coupled equations, one for the
two-point functions and one for the correlated parts. Physically,
the terms $\langle S_{i}^{\mu}\rangle G_{n,m}^{\rho,\beta}+\langle S_{n}^{\rho}\rangle G_{i,m}^{\mu,\beta}$
are similar to the Hartree and Fock contributions. They capture the
average effect of all the other spins on the two-point function, while
the correlated part $\mathcal{G}_{in,m}^{\mu\rho,\beta}$ contains
the corrections due to correlations with additional spins.

Our approximation to close the system of equations will assume that
correlated parts scale in some particular way with the parameters
of the system, such that they can be neglected when they are small
enough\cite{HierarchyCorr-Majorana,HIerarchy-Topology}. We will
organize terms in inverse powers of the coordination number $Z$,
where $Z$ generally denotes any coordination number required to describe
the model (for the Ising model just one coordination number is required,
while for the examples below we show that three different coordination
numbers are needed. In this case the smallest coordination number
will dominate the scaling). Doing this, it is possible to systematically
include higher order corrections by simply adding higher order correlated
parts.

It must be mentioned that the scaling of correlations converges more
slowly in 1D (it goes as $\sim Z^{-1}$, with $Z=2$). However, it
is still an interesting case, because in low dimensional systems quantum
corrections are usually enhanced. This facilitates the study of quantum
corrections due to the spin bath, and for this reason we will consider
the 1D case for the numerical results. However one must keep in mind
that the equations are general for arbitrary dimension. Furthermore,
we will show that including the correlated parts in the equation of
motion, allows to recover the exact quantum critical point in absence
of the spin bath, thus providing a good benchmark for our results.
For accurate results in low dimensional cases, the present technique
could also be combined with different fermionization or bosonization
techniques\cite{sachdev_2011}, and straightforwardly applying the
same Green's function analysis.

Let us apply this decoupling scheme to Eq.\ref{eq:EOM-G1} and Fourier
transform to frequency domain. One finds:\begin{widetext}
\begin{eqnarray}
\omega G_{n,m}^{\alpha,\beta} & = & \langle\left[S_{n}^{\alpha},O_{m}^{\beta}\right]_{\pm}\rangle+i\sum_{\mu}\epsilon_{\mu\alpha\rho}\left(B_{\mu}+\sum_{i\neq n}V_{i,n}^{\mu}\langle S_{i}^{\mu}\rangle-\sum_{l}A_{n,l}^{\mu}\langle I_{l}^{\mu}\rangle\right)G_{n,m}^{\rho,\beta}\label{eq:EOM-G2}\\
 &  & +i\sum_{\mu}\epsilon_{\mu\alpha\rho}\left(\sum_{i\neq n}V_{i,n}^{\mu}\left(\langle S_{n}^{\rho}\rangle G_{i,m}^{\mu,\beta}+\mathcal{G}_{in,m}^{\mu\rho,\beta}\right)-\sum_{l}A_{n,l}^{\mu}\left(\langle S_{n}^{\rho}\rangle Y_{l,m}^{\mu,\beta}+\mathcal{Y}_{ln,m}^{\mu\rho,\beta}\right)\right)\nonumber 
\end{eqnarray}
The first line corresponds to the uncorrelated contribution, while
the second line corresponds to the contribution from correlations
between spins. Notice the appearance of the Green's function $Y_{l,m}^{\mu,\beta}=-i\langle I_{l}^{\mu};O_{m}^{\beta}\rangle$,
which couples the equation of motion for the Ising spins with the
one for the bath spins:
\begin{eqnarray}
\omega Y_{n,m}^{\alpha,\beta} & = & \langle\left[I_{n}^{\alpha},O_{m}^{\beta}\right]_{\pm}\rangle+i\sum_{\mu}\epsilon_{\mu\alpha\rho}\left(\Delta_{\mu}-\sum_{i}A_{i,n}^{\mu}\langle S_{i}^{\mu}\rangle\right)Y_{n,m}^{\rho,\beta}\label{eq:EOM-Y1}\\
 &  & -i\sum_{\mu,i}\epsilon_{\mu\alpha\rho}A_{i,n}^{\mu}\left(\langle I_{n}^{\rho}\rangle G_{i,m}^{\mu,\beta}+\mathcal{Y}_{ni,m}^{\rho\mu,\beta}\right)\nonumber 
\end{eqnarray}
\end{widetext}Finally, one can write the equations in a more compact
form using matrix notation:
\begin{eqnarray}
\left(\omega-\hat{\mathbb{H}}\right)\cdot\hat{G} & = & \hat{\chi}^{\pm}+\hat{V}\cdot\hat{\mathcal{G}}\label{eq:EOM-Total}
\end{eqnarray}
where $\hat{G}$ is a vector containing all the system and bath two-point
functions, $\hat{\chi}^{\pm}$ contains the source terms coming from
the commutators/anti-commutators, $\hat{\mathcal{G}}$ contains all
extra contributions from correlations, $\hat{V}$ is the general interaction
matrix for the correlated parts, and $\hat{\mathbb{H}}$ is the effective
Hamiltonian for the two-point Green's functions. If the contribution
from the correlated parts $\hat{\mathcal{G}}$ is small enough, the
solution can be obtained by direct matrix inversion. This is a good
approximation near fixed points of the RG flow, where solutions are
approximately described by free spin Hamiltonians with renormalized
parameters. Furthermore, if one defines $\hat{g}=\left(\omega-\hat{\mathbb{H}}\right)^{-1}$,
the previous equation resembles a Dyson's equation, and from the correlated
parts it can be shown that $\hat{V}\cdot\hat{\mathcal{G}}$ can be
written as $\hat{\Sigma}\left(\omega\right)\cdot\hat{G}$, giving
rise to the self-energy and the familiar Dyson's equation.
\begin{equation}
\hat{G}=\hat{g}\cdot\hat{\chi}^{\pm}+\hat{g}\cdot\hat{\Sigma}\cdot\hat{G}\label{eq:Dyson}
\end{equation}

\section{Uncorrelated solution}

To close the system of equations one can neglect the correlated parts
in Eq.\ref{eq:EOM-Total}. Then the solution for the Green's function
is given by a matrix inversion. Eqs.\ref{eq:EOM-G2} and \ref{eq:EOM-Y1}
have two contributions from each interaction term, similar to the
well known Hartree and Fock terms. If one ignores the Fock contribution,
the coupling between Green's functions at different sites vanishes
and the equations are diagonal in real space. This is equivalent to
derive an effective-medium Hamiltonian for each spin, where all correlations
between sites are neglected. In this case one finds poles at $\omega_{s}\left(n\right)=\sqrt{\left(\sum_{\alpha}h_{s}^{\alpha}\left(n\right)\right)^{2}}$
for the central system, and at $\omega_{b}\left(n\right)=\sqrt{\left(\sum_{\alpha}h_{b}^{\alpha}\left(n\right)\right)^{2}}$
for the bath, where $h_{s}^{\mu}\left(n\right)=B_{\mu}+\sum_{i\neq n}V_{n,i}^{\mu}M_{i}^{\mu}-\sum_{l}A_{n,l}^{\mu}m_{l}^{\mu}$
and $h_{b}^{\mu}\left(n\right)=\Delta_{\mu}-\sum_{i}A_{i,n}^{\mu}M_{i}^{\mu}$.
As the Green's functions depend on the local magnetization $M_{n}^{\alpha}=\langle S_{n}^{\alpha}\rangle$
and $m_{n}^{\alpha}=\langle I_{n}^{\alpha}\rangle$, they need to
be determined self-consistency. Physically, this happens due to the
non-linearities introduced by the interaction term. Defining the spectral
function $J^{\alpha,\beta}\left(\omega\right)=i\left[g_{n,n}^{\alpha,\beta}\left(\omega+i\epsilon\right)-g_{n,n}^{\alpha,\beta}\left(\omega-i\epsilon\right)\right]\left(e^{\frac{\omega}{T}}\pm1\right)^{-1}$,
the self-consistency equations are obtained from the relation between
the statistical average and the spectral function $\frac{i}{2}\epsilon_{\mu\nu\alpha}M_{n}^{\alpha}=\int J^{\nu,\mu}\left(\omega\right)d\omega/2\pi$
\cite{Zubarev} (the $\pm$ sign corresponds to fermionic or bosonic
Green's function, respectively). The previous solution for the Green's
functions, in combination with the solutions of these non-linear equations
characterizing the local magnetization $M_{n}^{\alpha}$ and $m_{l}^{\alpha}$,
fully determine the properties of the system if correlations can be
neglected. Furthermore, as in this uncorrelated case the Green's functions
display simple poles only, one can rewrite the integral over frequency
as a sum over poles. Then the final form of the self-consistency equations
is: 
\begin{eqnarray}
M_{n}^{\alpha} & = & \frac{h_{s}^{\alpha}\left(n\right)}{2\omega_{s}\left(n\right)}\tanh\left(\frac{\omega_{s}\left(n\right)}{2T}\right)\label{eq:UncorrelatedSCE}\\
m_{n}^{\alpha} & = & \frac{h_{b}^{\alpha}\left(n\right)}{2\omega_{b}\left(n\right)}\tanh\left(\frac{\omega_{b}\left(n\right)}{2T}\right)
\end{eqnarray}
This result has been particularized for spin 1/2, but larger spins
can be studied similarly, as we discuss below for the specific case
of the Quantum Ising model. This simple result is a good first estimate
to see if the solutions are characterized by a magnetic texture. For
example, it can capture the anti-ferromagnetic ordering for the case
$V_{i,j}<0$. With this information one can build more complete solutions,
based on the symmetries of the ground states.

\paragraph*{Quantum Ising model longitudinally coupled to the spin bath:}

As a more specific case, let us consider the ferromagnetic Quantum
Ising model, longitudinally coupled to a bath of spins (we set $V_{i,j}^{x,y}=0$,
$B_{y,z}=0$, $A_{i,l}^{x,y}=0$ and initially $\vec{\Delta}=0$).
In this case the poles for the central system are $\omega_{s}=\sqrt{B^{2}+\left(V_{0}M_{z}-Z_{B}Am_{z}\right)^{2}}$,
while the ones for the bath are $\omega_{b}=Z_{BS}AM_{z}$. The self-consistency
equations can be directly obtained from Eq.\ref{eq:UncorrelatedSCE}
(to simplify notation we rename $V_{0}=Z_{S}V$, $V_{i,j}^{z}=V$
and $A=A_{i,l}^{z}$). We have assumed homogeneous magnetization to
simplify the expressions, which holds if the system has a uniform
magnetization solution. This not obvious in the presence of the spin
bath, as it can mediate interactions between the system spins, modulating
the initially spatially homogeneous solution. However, we assume bath
configurations which do not destroy the homogeneous magnetization.

When correlations are neglected, the physics is dominated by three
coordination numbers or connectivities: $Z_{S}$ corresponds to the
number of spins which directly interact in the central system, $Z_{B}$
to the number of bath spins connected to each spin, and $Z_{BS}$
to the number of spins directly interacting with each bath spin. The
coordination number $Z_{S}$ is well known in the analysis of the
Ising model, but with the addition of the bath of spins, now two additional
coordination numbers are needed. Interestingly, when correlations
are neglected, the results only depend on the topology of the lattice
(i.e., the graph connectivity), while the geometrical effects will
appear when correlations between sites are added.

An obvious question now is how the different phase transitions are
affected by the bath. To obtain the Curie temperature, $T_{c}$ we
fix $B=0$ and expand to third order in powers of $M_{z}$. The solution
for $M_{z}=0$ is found to be: 
\begin{eqnarray}
T_{c} & \simeq & \frac{V_{0}}{8}+\frac{\sqrt{\left(\frac{V_{0}}{2}\right)^{2}+Z_{B}Z_{BS}A^{2}\frac{1+8\left|\vec{P}\right|+4\left|\vec{P}\right|^{2}}{6}}}{4}\label{eq:Curie-Temp}
\end{eqnarray}
It is easy to see that $T_{c}\left(A\rightarrow0\right)=V_{0}/4$,
in agreement with the Curie-Weiss law for the Ising model. The critical
temperature seems to increase with all coordination numbers and with
the total spin of the bath $\left|\vec{P}\right|$ (see details of
the calculation in the Appendix\ref{sec:Appendix1}, where arbitrary
spin value for the bath is assumed). This result requires some clarification,
as it is known that the $T$-dependence for the longitudinal magnetization
should not be affected if the spin couples to a single bath spin\cite{Page_1984}.
The reason for this discrepancy is the lack of correlations between
the Ising spin and the local bath spin for the uncorrelated solution.
If one considers instead the electro-nuclear basis at each site, or
includes correlations between the bath and the Ising spin, both results
agree. To confirm this, we have calculated in the Appendix the deviation
from the exact solution for the case of a single Ising spin coupled
to a bath spin. It shows that the limits of high and low temperature
are well captured by the uncorrelated solution. However, as one goes
to intermediate temperatures, correlations between the two spins become
important, and the exact solution deviates from the uncorrelated one
(Fig.\ref{fig:Comparison} in the Appendix). This provides a practical
example of the importance of correlations, specially near the phase
transitions, and also confirms the validity of our results near the
fixed points of the theory.

Importantly, our solution can describe the case in which the spin
bath is non-local ($Z_{BS}>1$), which has not been previously discussed.
Also, as previously indicated, the electro-nuclear correlations can
be easily incorporated by exactly diagonalizing the local electro-nuclear
Hamiltonian \cite{Ryan-Spinbath}.

\begin{figure}
\includegraphics[scale=0.45]{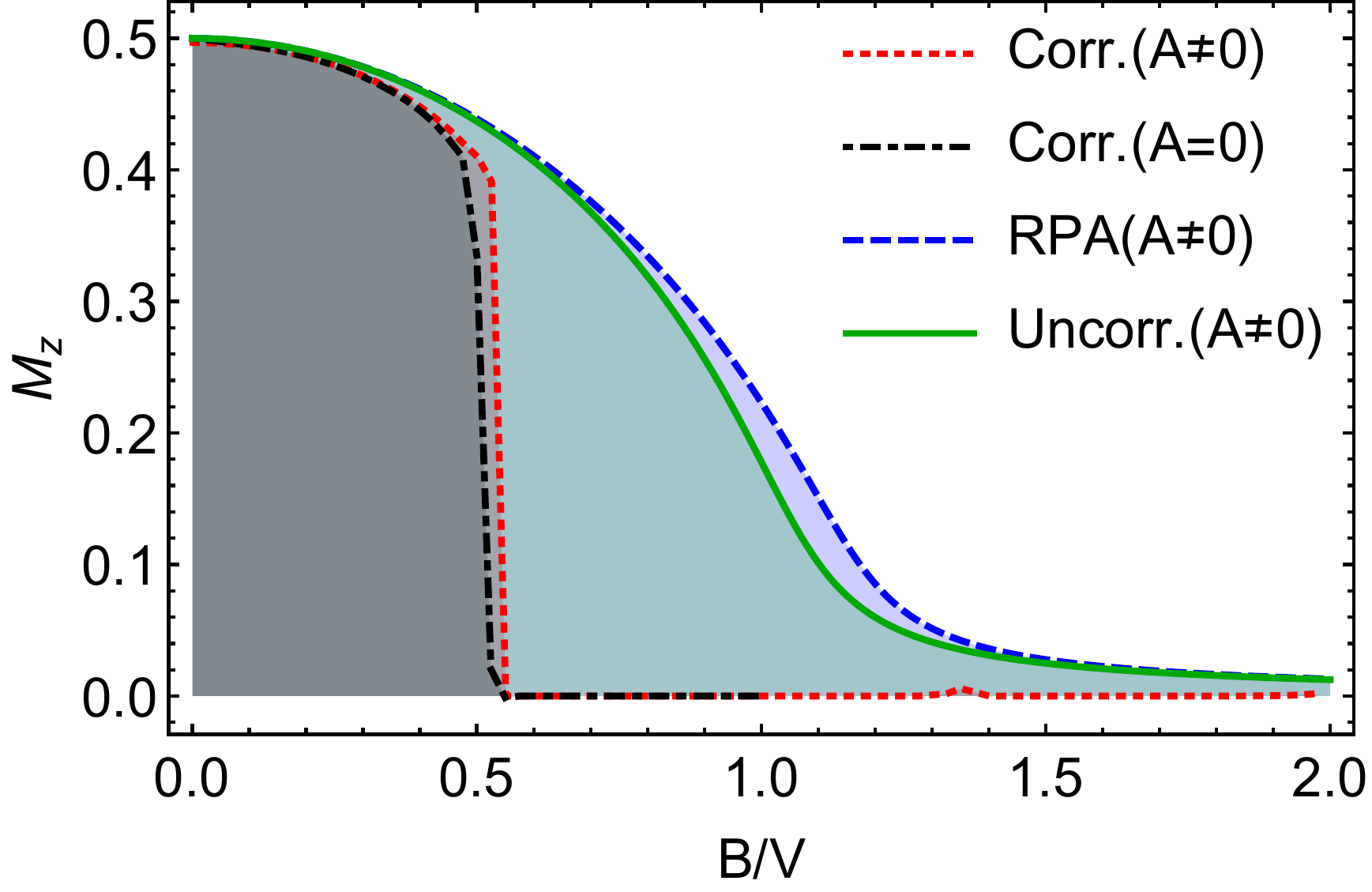}

\caption{\label{fig:Phase-Diag1}Phase diagram for the uncorrelated solution
as a function of $B/V$. (Black) Isolated Ising model at low T. (Red)
Ising model longitudinally coupled to a static bath for $A/V=0.05$
and low $T$. (Blue) Ising model longitudinally coupled to a static
bath at higher $T$. Temperature disorders the spin bath and unblocks
the phase transition. (Green) Ising model at low $T$ longitudinally
coupled to a dynamical spin bath ($\Delta\protect\neq0$), which unblocks
the QPT as well. We have considered $Z_{BS}=Z_{B}=1$ for the plot.}
\end{figure}

To find the critical field $B_{c}$ which characterizes the quantum
phase transition (QPT) we take the limit $T\rightarrow0$ and expand
to first order in powers of $M_{z}$:
\begin{eqnarray}
M_{z} & \simeq & \frac{Z_{B}A\left|\vec{P}\right|}{2\sqrt{B^{2}+A^{2}\left|\vec{P}\right|^{2}Z_{B}^{2}}}+\frac{B^{2}V_{0}M_{z}}{2\left[B^{2}+A^{2}\left|\vec{P}\right|^{2}Z_{B}^{2}\right]^{3/2}}\label{eq:RemnantMz}
\end{eqnarray}
This shows that $M_{z}\left(T\rightarrow0\right)=0$ is not a solution,
and the QPT between the ferromagnetic and the paramagnetic phases
is blocked due to a remnant magnetization induced by the bath. Furthermore,
this remnant magnetization scales as the first term in Eq.\ref{eq:RemnantMz}.
This means, strictly speaking, that there is no QPT, although transverse
terms in the interaction can recover the QCP. Interestingly, all three
coordination numbers appear in the result for the critical temperature,
while $Z_{BS}$ is absent in the expression for the quantum critical
point. Fig.\ref{fig:Phase-Diag1} compares the characteristic behavior
of the longitudinal magnetization in the presence of the spin bath
for different cases. It shows that at low temperatures, the longitudinal
coupling to the bath blocks the QPT (red), but it can be recovered
by adding a transverse field $\Delta_{x}\neq0$ to the bath spins
(green). Increasing the temperature also unblocks the transition (blue),
because it disorders the bath, but the phase transition is not strictly
at $T=0$.

The phase diagram in absence of correlations captures qualitatively
the different phases, but the critical point is overestimated. For
the 1D Ising model it is well known that the exact critical field
corresponds to $B/V=1/2$ , and this discrepancy happens due to corrections
produced by correlations. In the last section we will show that a
very good agreement is obtained if correlated parts are added, but
let us first consider the effect of the Fock terms in the $T=0$ regime.

\section{Correlations between spins}

In general, quantum systems are made of interacting particles, and
although their interactions may be weak, the correlations generated
by these interactions may still have important consequences. For example,
it is well known that near classical and quantum phase transitions,
systems display macroscopic correlations, at length scales much larger
than the typical length scales present in the microscopic model. This
is a particular case of emergence, where the system behaves in a different
way than the particles in the underlying microscopic model. Therefore,
it is clear that in some cases correlations are important, and results
where correlations are neglected will be significantly affected. Another
important case, present in low dimensional quantum systems, is when
due to confinement, the role of quantum fluctuations is enhanced.
A well known consequence of this is the absence of certain phase transitions
in low dimensional models.

To include the effect of correlations one just needs to keep the terms
that were previously neglected in the equation of motion. The simplest
contribution is the Fock term, which couples two-point Green's functions
at different sites. With this term added, two-point correlations are
captured. If the system is translationally invariant, the Green's
function is diagonal in $k$-space and the matrix inversion can be
performed analytically. The main effect of the Fock term is to make
quasiparticles dispersive, and to interpolate between the fixed points
of the theory, where correlations between sites can be neglected (e.g.,
between the $B=0$ and the $V=0$ limits of the Quantum Ising model).
For the present case with a spin bath, this correction can also correlate
the Ising and the bath Green's functions (see terms $\langle S_{n}^{\rho}\rangle Y_{l,m}^{\mu,\beta}$
and $\langle I_{n}^{\rho}\rangle G_{i,m}^{\mu,\beta}$ in Eq.\ref{eq:EOM-G2}
and Eq.\ref{eq:EOM-Y1}, respectively). This implies that in general,
magnons in the Ising model become a mixture of Ising magnons and excitations
in the spin bath. Finally, the correction due to the correlated part
of the three-point function requires to calculate a new equation of
motion, which in the present case will also encode magnon-magnon interactions.
Formally, it can be shown that the equation of motion for the correlated
part of the three-point function can be written in the next form:
\begin{equation}
\left(\omega-\mathbb{\hat{H}}_{\mathcal{G}}\right)\cdot\hat{\mathcal{G}}=\hat{\Lambda}_{0}+\hat{\Lambda}\cdot\hat{G}\label{eq:CorrelatedEOM}
\end{equation}
once it is truncated by neglecting four-point correlations, which
are expected to scale as $Z^{-2}$\footnote{In the present case the system is characterized by three different
coordination numbers, and the smallest one will dominate this scaling.}. In Eq.\ref{eq:CorrelatedEOM} we have defined $\mathbb{H}_{\mathcal{G}}$
as the ``Hamiltonian'' for the correlated part, which is obtained
from its equation of motion, and $\hat{\Lambda}$ and $\hat{\Lambda}_{0}$
are the two types of source terms which can be present. The solution
can be written as: 
\begin{equation}
\hat{\mathcal{G}}=\frac{\hat{\Lambda}_{0}+\hat{\Lambda}\cdot\hat{G}}{\omega-\hat{\mathbb{H}}_{\mathcal{G}}}
\end{equation}
 Inserting this result in Eq.\ref{eq:EOM-Total}, yields the solution
for the two-point function including correlations from the three-point
function:
\begin{eqnarray}
\hat{G} & = & \frac{\hat{\chi}^{\pm}+\hat{\Sigma}_{0}}{\omega-\hat{\mathbb{H}}-\hat{\Sigma}}\label{eq:formal-Sol-G}
\end{eqnarray}
where the self-energies are defined as $\hat{\Sigma}\equiv\frac{\hat{V}\cdot\hat{\Lambda}}{\omega-\mathbb{H}_{\mathcal{G}}}$
and $\hat{\Sigma}_{0}\equiv\frac{\hat{V}\cdot\hat{\Lambda}_{0}}{\omega-\mathbb{H}_{\mathcal{G}}}$.
Equation \ref{eq:formal-Sol-G} shows that in general, $\hat{\Sigma}$
modifies the pole structure of the Green's function , while $\hat{\Sigma}_{0}$
modifies the spectral weight.

Although the formal description of the previous solutions seems quite
simple, a full solution can be a challenging task due to the self-consistency
equations, and in general, one needs to make use of numerical methods.
Nevertheless, it is also possible to obtain some analytical results
by means of perturbative expansions and other approximation methods.
In this work we will focus on the effect of two-spin correlations,
which capture the magnon quasiparticles. The specific effect of the
correlated parts will be analyzed only for the correction to the longitudinal
magnetization, while more general effects will be discussed in future
works.

\paragraph*{Correlations for the Quantum Ising model longitudinally coupled to
the spin bath:}

\begin{figure}
\includegraphics[scale=0.45]{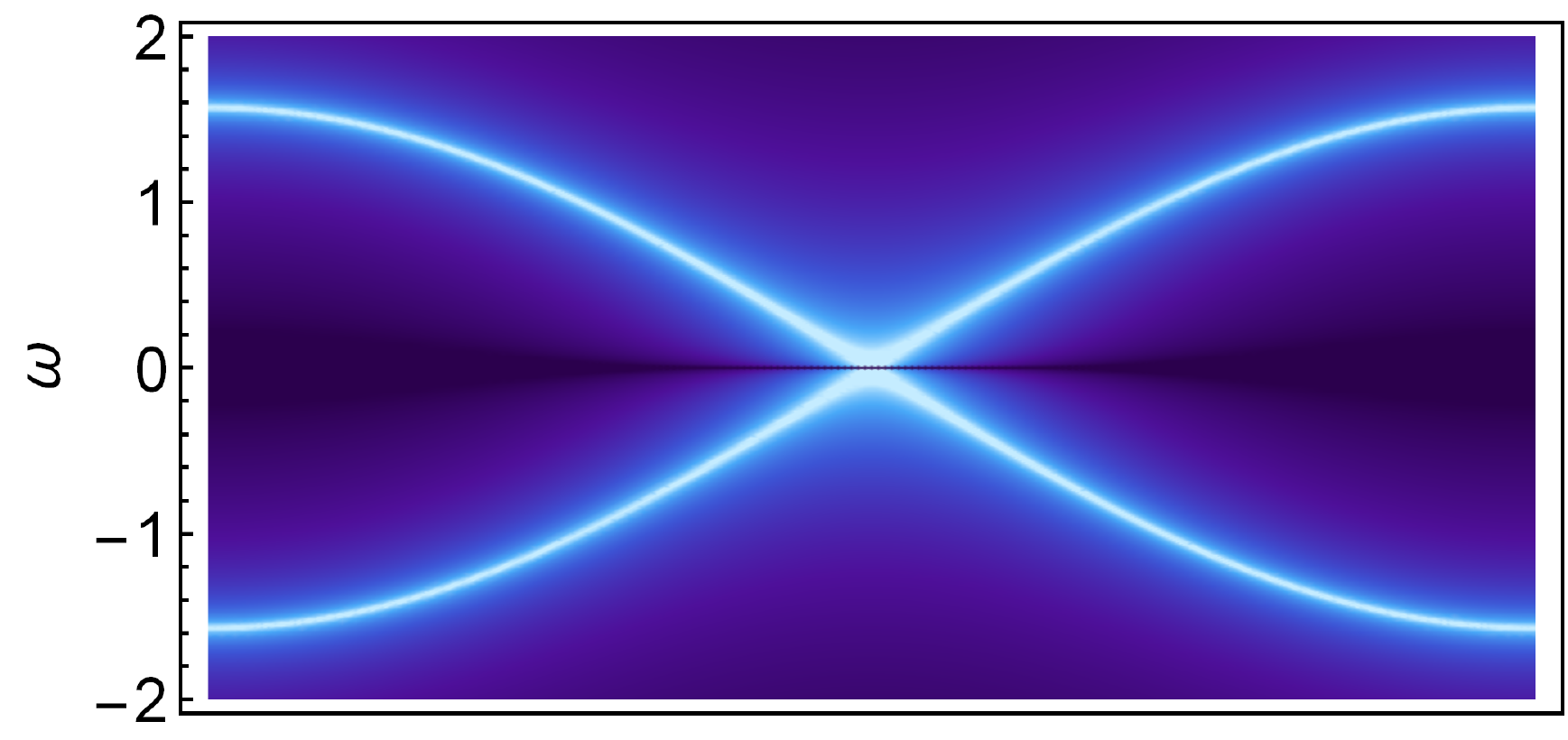}

\includegraphics[scale=0.45]{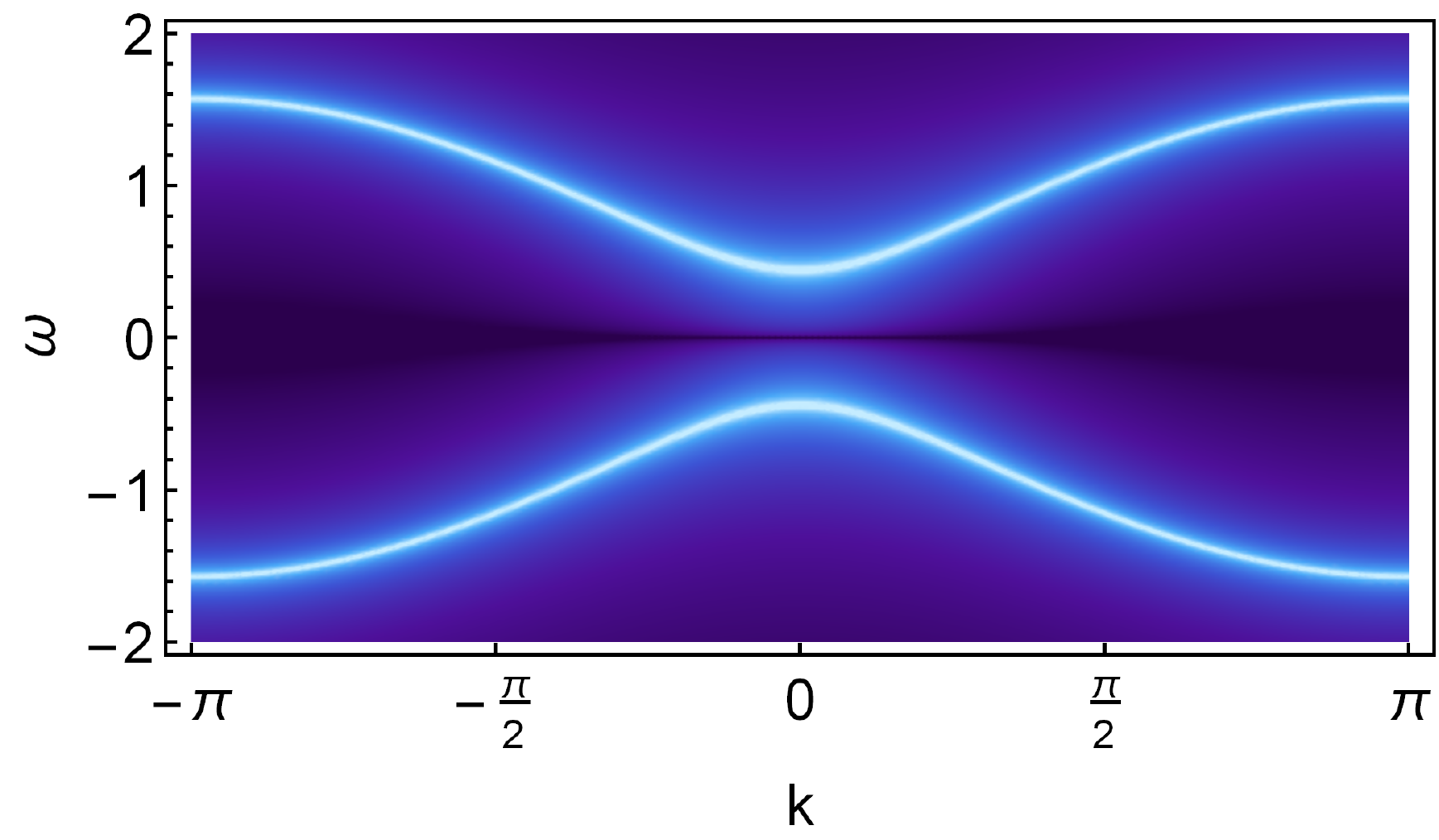}

\caption{\label{fig:SpectralFunction}Complex part of the bosonic Green's function
$G_{k,k}^{z,z}\left(\omega\right)$ as a function of $k$ and $\omega$.
(Top) In absence of the spin bath, the magnon gap closes at the critical
point $B_{c}\simeq1.1V$. (Bottom) Gap does not close at the critical
point if the longitudinal coupling to the bath is included. We have
considered $A=0.05V$ and a 1D lattice with $Z_{B}=Z_{BS}=1$. The
results are expressed in logarithmic scale to enhance the contrast.}

\end{figure}

\begin{figure}
\includegraphics[scale=0.78]{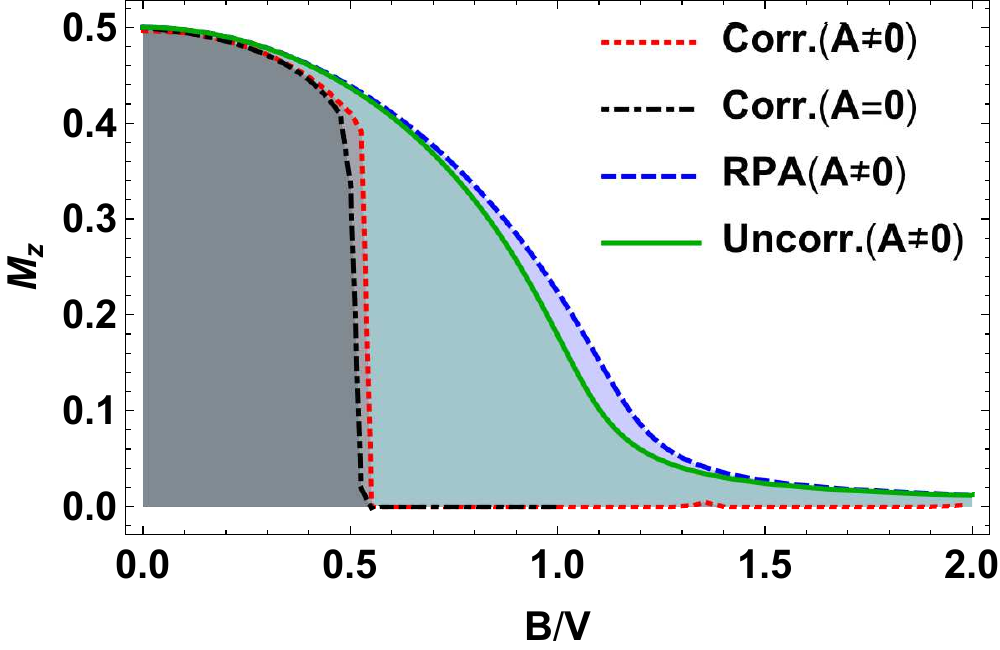}

\caption{\label{fig:Magnetization-Comparison}$M_{z}$ vs $B/V$ for the Ising
model longitudinally coupled to a bath of spins at $T=0$. (Green)
Uncorrelated solution for $A=0.05V$ and $Z_{B}=1$. (Blue) RPA solution
including for $A=0.05V$ and $Z_{B}=1$. (Black) Solution including
correlated parts without the spin bath ($A=0$). (Red) Solution including
correlations for a large spin bath ($A=0.2V$ and $Z_{B}=100$).}

\end{figure}

\begin{figure}
\includegraphics[scale=0.5]{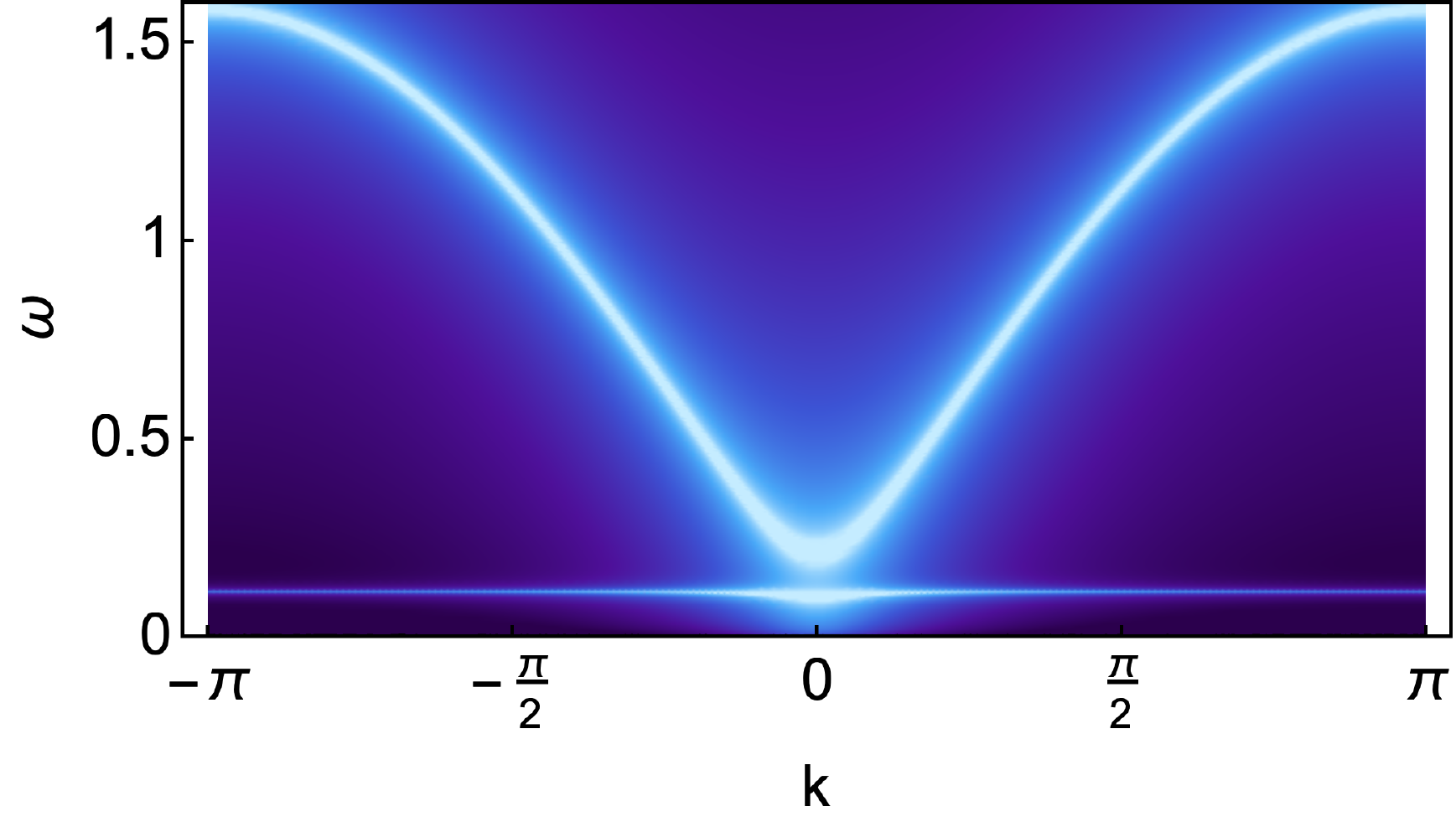}

\caption{\label{fig:Electro-nuclear-mode}Emergence of an electro-nuclear mode
at the critical point when system and bath get correlated due to a
finite $\Delta=10^{-3}B$.}
\end{figure}

Now we include correlations in the transverse Ising model, as the
significance of this model for current quantum computing architectures
is crucial. The addition of the Fock term modifies the previous equations
of motion, and can couple the Ising and bath spins. For the case with
$\vec{\Delta}=0$ the main difference with the previous solution is
the change in the quasiparticle excitations, from localized spin flips
to magnons with dispersion relation:
\begin{equation}
\omega_{k}=\sqrt{\left(M_{z}V_{0}-Z_{B}Am_{z}\right)^{2}+B^{2}-BM_{x}V_{k}}\label{eq:Magnons-dispersion}
\end{equation}
Notice that the last term vanishes at both fixed points $B=0$ and
$V=0$, agreeing with the uncorrelated solutions. However, as one
interpolates between the two, the quasiparticles turn mobile and the
band structure becomes dispersive. Results including the Hartree and
Fock terms in the equations of motion are similar to the Random Phase
Approximation (RPA)\citep{Ryan-Spinbath}. The self-consistency equations
are now:
\begin{eqnarray}
M_{z} & = & \frac{1}{2}\frac{M_{z}V_{0}-Z_{B}Am_{z}}{\frac{1}{N}\sum_{k}\omega_{k}\coth\left(\frac{\omega_{k}}{2T}\right)}\\
M_{x} & = & \frac{1}{2}\frac{B}{\frac{1}{N}\sum_{k}\omega_{k}\coth\left(\frac{\omega_{k}}{2T}\right)}
\end{eqnarray}
In absence of the spin bath ($A=0$), the phase diagram does not change
much with respect to the uncorrelated result. The critical field $B_{c}$
shifts towards slightly larger values, stabilizing the ferromagnetic
phase. On the other hand, this approximation captures that, at the
quantum phase transition, the magnon gap closes, as it can be seen
in the spin-spin correlation function (Fig.\ref{fig:SpectralFunction},
top).
When the spin bath is included ($A\neq0$), the difference between
the uncorrelated and correlated phase diagrams gets reduced (see Fig.\ref{fig:Magnetization-Comparison},
green and blue lines). The reason is that two-point correlations between
Ising spins dominate near the unperturbed Ising critical point, which
is now shifted. This can be seen in Fig.\ref{fig:Magnetization-Comparison},
where the main differences happen near the Ising critical point $B_{c}\simeq1.1V$,
and the two solutions match again for larger $B$. Therefore, spin-spin
correlations do not change quantitatively the phase diagram, in the
presence of the spin bath.
On the other hand, the effect of the bath in the correlation functions
is more crucial, as it leads to a mode softening at the critical point.
Without the spin bath, the magnon gap closes, signaling a divergence
of the correlation length between Ising spins; however, the presence
of the remnant field produced by the spin bath makes the gap finite
for all $B$, and the correlation length between spins remains finite.
Furthermore, the effect of a dynamical bath due to a transverse field
$\Delta\neq0$ modifies this picture non-trivially. The fact that
Eq.\ref{eq:Magnons-dispersion} corresponds to Ising magnons under
an effective longitudinal field $\sim Z_{B}Am_{z}$ is not a coincidence.
When the bath is static ($\Delta=0$), system and bath do not get
entangled, but making the bath dynamical ($\Delta\neq0$) correlates
both systems and modifies the quasiparticle picture. Fig.\ref{fig:Electro-nuclear-mode}
shows the appearance of an electro-nuclear mode with non-vanishing
spectral weight. This mode emerges near $\omega\simeq0$ when $\kappa\equiv Z_{B}Z_{BS}A^{2}B\Delta m_{x}M_{x}\neq0$
and corresponds to a mixture of magnon and bath excitation, as it
can be seen from the poles of the Green's function:
\begin{equation}
\tilde{\omega}_{k}=\pm\sqrt{\frac{\Omega^{2}+\omega_{k}^{2}}{2}\pm\sqrt{\kappa+\left(\frac{\Omega^{2}-\omega_{k}^{2}}{2}\right)^{2}}}
\end{equation}
where $\Omega=\sqrt{\Delta^{2}+Z_{BS}^{2}A^{2}}$ (details in the
Appendix). The presence of this mode has been observed in \citep{LiHoFPhaseTransition}
and discussed in detail in ref.\citep{Ryan-Spinbath}. There it is
shown that this mode does not carry spectral weight unless the bath
is dynamical, and that it allows to recover the QCP. 
Finally, we discuss the impact of adding the correlated parts in the
equation of motion. As previously mentioned, their effect can be incorporated
in terms of a self-energy, but in the general case, the self-energy
will be a complicated function of frequency and momentum. To estimate
their effect we analyze the longitudinal magnetization, which can
be addressed more easily by using the Majorana representation for
spins\citep{MajoranaRep}:
\begin{equation}
S^{\alpha}=-\frac{i}{2}\epsilon_{\alpha\theta_{1}\theta_{2}}\eta^{\theta_{1}}\eta^{\theta_{2}}
\end{equation}
This representation simplifies the self-consistency equations, and
makes the self-energy for the longitudinal magnetization $\Sigma^{zz}\left(\vec{k},\omega\right)$
a function of $\omega$ only. Furthermore, unlike the Wigner-Jordan
transformation, this representation is local and the separation between
correlated and uncorrelated parts is analogous to the spins case.
The derivation of the equations of motion, as well as the decoupling
scheme is analogous to the case of the spin representation, as described
in the Appendix.

The phase diagram including the correlated parts is shown in Fig.\ref{fig:Magnetization-Comparison}.
In absence of the spin bath ($A=0$, black line), the ferromagnetic
phase shrinks and the QCP shifts to $B/V\simeq1/2$, which is the
exact result for the 1D Ising model. When the spin bath is included
($A\neq0$, red line), the phase boundary shifts towards larger values
of $B/V$, as it happened in the uncorrelated phase diagram. However,
in this case the remnant magnetic field produced by the spin bath
is highly renormalized by the correlated parts. This can be seen from
the specific values chosen for the bath in Fig.\ref{fig:Magnetization-Comparison}
($A=0.2V$ and $Z_{B}=100$), which for the uncorrelated solution
would produce a phase boundary shifted towards much larger values
of $B/V$. Importantly, although the QPT seems to be recovered when
correlated parts are included, this is not the case. One can check
that $M_{z}\left(T=0\right)=0$ is not a fixed point of the self-consistency
equations, although its final value is very small due to the large
renormalization.

\paragraph*{Conclusions:}

We have shown that a large variety of spin models interacting with
spin baths can be treated using the equation of motion technique.
This approach allows for non-perturbative solutions, even to lowest
order, and using our decoupling scheme one can include corrections
in a systematic way. Neglecting correlations yields the standard MF
solution, while adding the lowest order correlations between pairs
of spins reproduces the RPA results.

As a concrete example, we have analyzed the Quantum Ising model coupled
to a local bath of spins, where each Ising spin is coupled to a set
of $Z_{B}$ independent spins. This case is interesting due to the
changes produced by the longitudinal interaction on the critical properties.
In the absence of correlations we have obtained solutions for arbitrary
spin value in the bath, and demonstrated that a remnant magnetic field
produced by the bath can block the quantum phase transition. When
correlations are included, system and bath quasiparticles can mix,
transforming the standard magnons into electro-nuclear modes. We have
shown that the effect of internal dynamics in the spin bath entangles
the system and bath spins, and that this leads to the emergence of
electro-nuclear modes near the phase transition.

When higher correlations are added (i.e., the correlated parts of
the Green's functions), some quantities are accurately captured. For
the Quantum Ising model we have obtained the exact critical point
in 1D and shown that the effect of the spin bath gets highly renormalized
by the virtual processes.

Future directions for this work include the study of more complicated
spin models, as for example with transverse terms and dipolar interactions,
or considering more complex connectivity between bath and system (e.g.,
if a bath spin can interact with more than a single Ising spin, the
results would appreciably change). To conclude, effective models of
spin systems interacting with localized modes are ubiquitous in experiments
at low temperature. Some examples are: dangling bonds \cite{DanglingBonds},
nuclear spins \cite{LiHoFPhaseTransition,MorelloNuclearSpinBathSingleMolecule,Nuclearspinbathquantumdots},
paramagnetic impurities \cite{ParamagneticImpurities,ParamagneticImpuritiesII},
and localized vibrational modes \cite{Discretphononmodes}. Therefore
a a theoretical approach that allows to systematically study them
is very relevant for the field.

We would like to thank R. D. McKenzie for helpful discussions. This
work was supported by the National Scientific and Engineering Research
Council of Canada and A.G-L. acknowledges the Juan de la Cierva program.

\bibliographystyle{phaip}
\bibliography{bibliografia-Spinbath}

\newpage\begin{widetext}

\appendix

\section{Comparison between the uncorrelated and the exact solution for two
spins\label{sec:Appendix1}}

Here we discuss the limitations of the uncorrelated solution in terms
of a two spin system. Consider the next Hamiltonian:
\begin{equation}
H=-B_{x}S^{x}-B_{z}S^{z}-\Delta_{z}I^{z}+AI^{z}S^{z}
\end{equation}
where a single spin is coupled to a single bath spin. The interaction
between the two spins is longitudinal and proportional to $A$. In
addition, the Ising spin has a longitudinal field $B_{z}$ which mimics
the effect of the field produced by all the other Ising spins in the
Ising model, and a transverse field $B_{x}$. Also the bath spin can
be biased by a longitudinal field $\Delta_{z}$. The exact calculation
is possible due to the small size of the Hilbert space, and the statistical
average for the magnetization of the Ising spin can be obtained analytically:
\begin{eqnarray}
M_{z}=Z^{-1}\textrm{Tr}\left\{ e^{-\frac{H}{T}}S^{z}\right\}  & = & e^{-\frac{\Delta}{2T}}\left(A+2B_{z}\right)\frac{\sinh\left(\frac{\sqrt{\left(A+2B_{z}\right)^{2}+4B^{2}}}{4T}\right)}{Z\sqrt{\left(A+2B_{z}\right)^{2}+4B^{2}}}-e^{\frac{\Delta}{2T}}\left(A-2B_{z}\right)\frac{\sinh\left(\frac{\sqrt{\left(A-2B_{z}\right)^{2}+4B^{2}}}{4T}\right)}{Z\sqrt{\left(A-2B_{z}\right)^{2}+4B^{2}}}
\end{eqnarray}
where
\begin{equation}
Z=2e^{\frac{\Delta}{2T}}\cosh\left(\frac{\sqrt{\left(A-2B_{z}\right)^{2}+4B_{x}^{2}}}{4T}\right)+2e^{-\frac{\Delta}{2T}}\cosh\left(\frac{\sqrt{\left(A+2B_{z}\right)^{2}+4B_{x}^{2}}}{4T}\right)
\end{equation}
is the partition function. On the other hand, we make use of the self-consistency
equations for the magnetization, obtained from the uncorrelated solutions
of the Green's functions in the main text:
\begin{eqnarray}
M_{z} & = & \frac{B_{z}-Am_{z}}{2\sqrt{B_{x}^{2}+\left(B_{z}-Am_{z}\right)^{2}}}\tanh\left(\frac{\sqrt{B_{x}^{2}+\left(B_{z}-Am_{z}\right)^{2}}}{2T}\right)\\
M_{x} & = & \frac{B_{x}}{2\sqrt{B_{x}^{2}+\left(B_{z}-Am_{z}\right)^{2}}}\tanh\left(\frac{\sqrt{B_{x}^{2}+\left(B_{z}-Am_{z}\right)^{2}}}{2T}\right)\\
m_{z} & = & \frac{\Delta_{z}-AM_{z}}{2\sqrt{\left(\Delta_{z}-AM_{z}\right)^{2}}}\tanh\left(\frac{\sqrt{\left(\Delta_{z}-AM_{z}\right)^{2}}}{2T}\right)
\end{eqnarray}
The comparison between the two solutions is shown in Fig.\ref{fig:Comparison}.
It shows that in the absence of a transverse field ($B_{x}\simeq0$)
the longitudinal bath does not affect the Curie temperature, which
is missed by the uncorrelated solution. On the other hand, for non-vanishing
transverse field (i.e., when the Ising model becomes ``quantum''),
the bath modifies the behavior with temperature, specially in the
low temperature regime.

Notice that the exact solution in terms of Green's functions is also
possible in this case, by just including the correlated part. This
indicates that the intermediate temperature regime needs to be characterized
including correlations (as one would expect, specially at the Curie
temperature, where the correlation length diverges). 
\begin{figure}[H]
\includegraphics[scale=0.6]{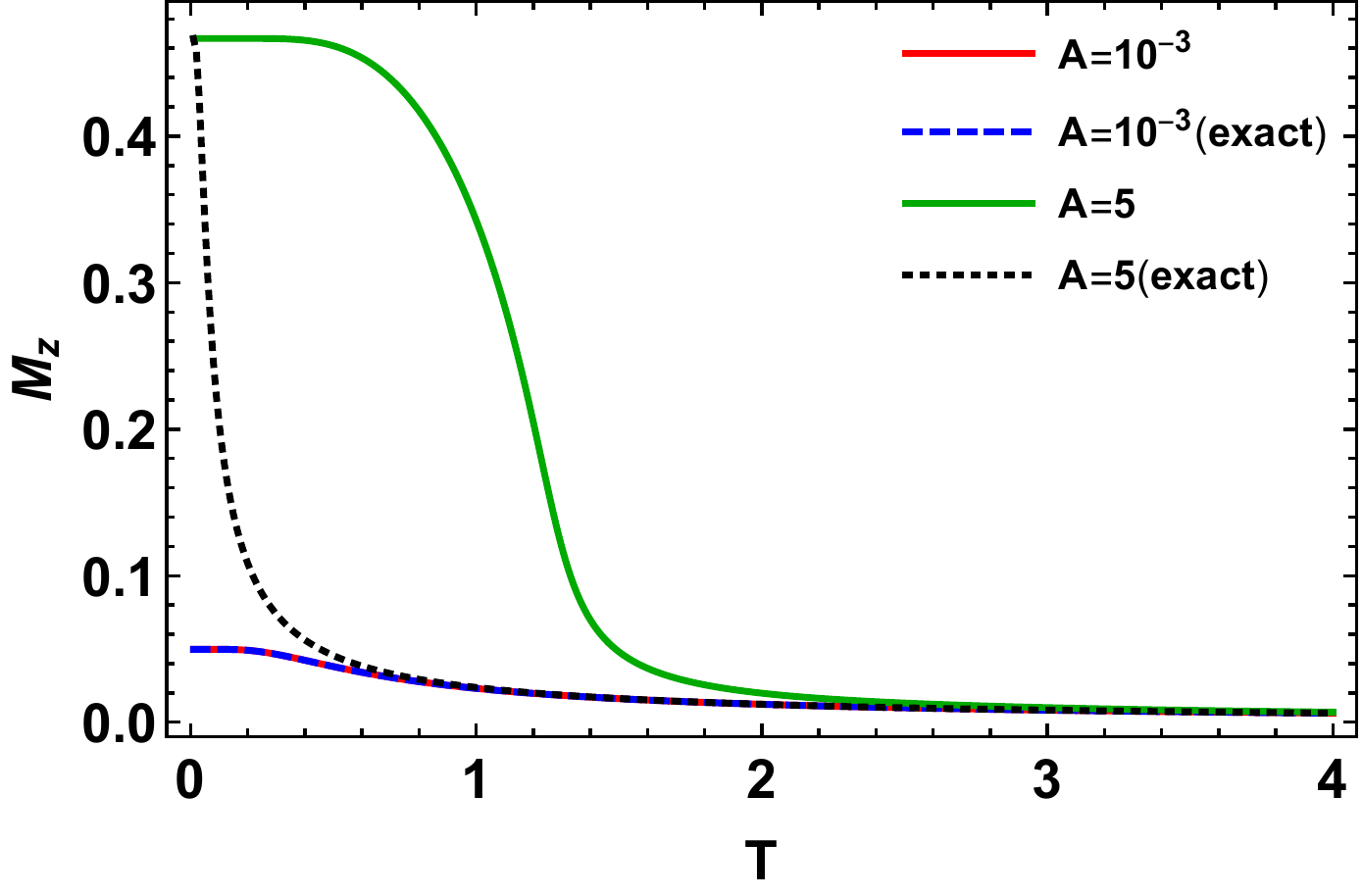}\includegraphics[scale=0.6]{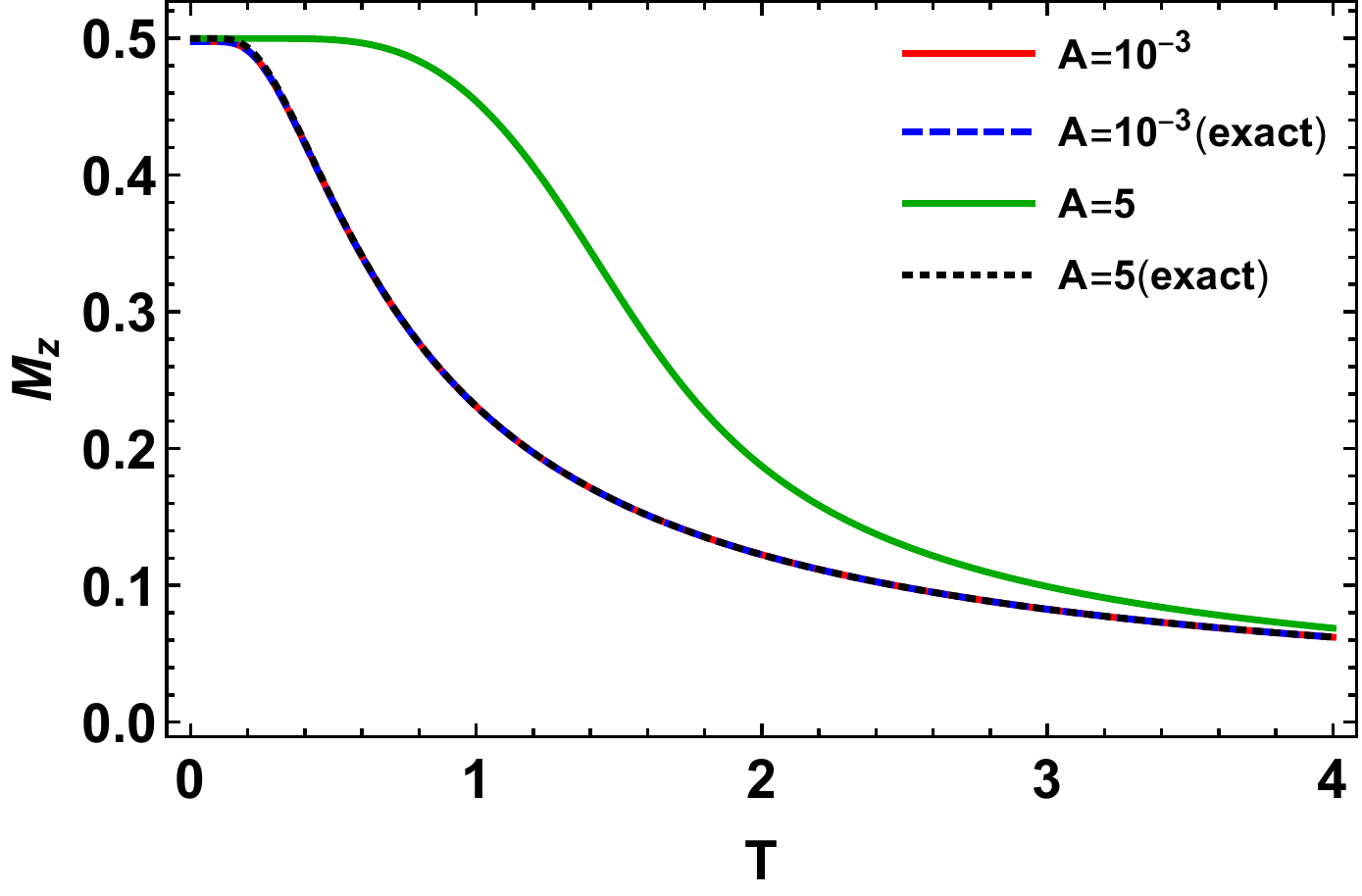}

\caption{\label{fig:Comparison}Comparison between the self-consistent solution
and the exact calculation. (Left) Case with $B_{z}=0.1$ and $B_{x}=1$,
and (right) case with $B_{z}=1$ and $B_{x}=0.1$. In general, the
uncorrelated solution and the exact one, perfectly agree at low and
high temperature, while for intermediate values they differ (for weak
$A$ they agree very well in general). When the system is in the analog
of a FM phase, with $B_{x}\ll B_{z}$ (right), the hyperfine coupling
(even for large $A$) would not make an important change in the Curie
temperature (which happens for intermediate values of $T$). This
is in contrast with the case $B_{x}\gg B_{z}$ (left), where specially
at low $T$ the cases with small and large $A$ are quite different.}
\end{figure}

\section{Quantum Ising model coupled to a spin bath}

The case of the Quantum Ising model is obtained from the general equation
of motion by setting $B_{y,z}=0$, $V_{i,j}^{x,y}=0$, $\Delta_{x,y}=0$
and $A_{i,l}^{x,y}=0$. Furthermore, if the bath spins are static
($\vec{\Delta}=0$), it is useful to consider for its basis the set
of eigenstates $|P,P_{z}\rangle$, where $P$ labels the spin value
$P\in\left[\left|\vec{P}\right|,\left|\vec{P}\right|-1,\ldots,1/2\right]$
and $P_{z}\in\left[P,P-1,\ldots,-P\right]$ its projection. Their
equation of motion is calculated analogously, from the Heisenberg
equation:
\begin{equation}
\partial_{t}\hat{O}=i\left[H,\hat{O}\right]
\end{equation}
The advantage of using this basis is that it allows to calculate the
solutions for arbitrary spin value. The Hamiltonian reads in this
basis:
\begin{equation}
H_{S}=-B\sum_{i}S_{i}^{x}-\sum_{i,j>i}V_{i,j}S_{i}^{z}S_{j}^{z}-\Delta_{z}\sum_{l}\sum_{P,P_{z}}P_{z}X_{P,l}^{P_{z},P_{z}}+\sum_{i,l}A_{i,l}S_{i}^{z}\sum_{P,P_{z}}P_{z}X_{P,l}^{P_{z},P_{z}}
\end{equation}
where 
\[
X_{P,l}^{P_{z},P_{z}}=|P,P_{z}\rangle_{l}\langle P,P_{z}|_{l}
\]
is the projector onto the eigenstate of the bath at site $l$. Neglecting
correlations one finds that the average value of each projector, and
the magnetization $m_{n}^{z}=\sum_{P_{z},P}P_{z}\langle X_{P,n}^{P_{z},P_{z}}\rangle$
are given by:
\begin{eqnarray}
\langle X_{P,n}^{a,a}\rangle & = & \frac{e^{\left(a-1\right)\frac{h_{b}}{T}}}{2}\frac{\left(e^{\frac{h_{b}}{T}}-1\right)^{2}}{\cosh\left[\frac{h_{b}}{T}\left(\left|\vec{P}\right|+1\right)\right]-\cosh\left(\frac{h_{b}}{2T}\right)}\\
m_{n}^{z} & = & -\frac{1}{4}\text{csch}\left(\frac{h_{b}}{2T}\right)\frac{3+\cosh\left(\frac{h_{b}}{T}\right)-2\left(2+\left|\vec{P}\right|\right)\cosh\left[\frac{h_{b}}{2T}\left(1+2\left|\vec{P}\right|\right)\right]+2\left|\vec{P}\right|\cosh\left[\frac{h_{b}}{2T}\left(3+2\left|\vec{P}\right|\right)\right]}{\cosh\left(\frac{h_{b}}{2T}\right)-\cosh\left[\frac{h_{b}}{T}\left(\left|\vec{P}\right|+1\right)\right]}
\end{eqnarray}
where $h_{b}\left(n\right)=\Delta_{z}-\sum_{i}A_{i,n}M_{i}^{z}$.
This solution is specific for half-integer spins, but the case for
integer spins can be calculated analogously. Finally, we use the solution
for the fermionic Green's function to obtain the self-consistency
equations for the magnetization:
\begin{eqnarray}
M_{z} & = & \frac{h_{s}}{2\omega_{s}}\tanh\left(\frac{\omega_{s}}{2T}\right)\\
M_{x} & = & \frac{B}{2\omega_{s}}\tanh\left(\frac{\omega_{s}}{2T}\right)
\end{eqnarray}
With these results one can now estimate the Curie temperature and
the critical field. For the Curie temperature we are looking for solutions
with $B=0$. We consider the case where the bath is unbiased by an
external field to simplify the expressions ($\Delta_{z}=0$). Expanding
for small $M_{z}$ up to third order, and solving for the value of
$T$ that makes $M_{z}=0$, we find:
\begin{eqnarray}
T_{c} & = & \frac{V_{0}}{8}+\frac{1}{4}\sqrt{\left(\frac{V_{0}}{2}\right)^{2}+Z_{B}Z_{BS}A^{2}\frac{1+8\left|\vec{P}\right|+4\left|\vec{P}\right|^{2}}{6}}
\end{eqnarray}
It shows that for this approximation, the Curie temperature would
increase with all the coordination numbers (remember $V_{0}=Z_{s}V$),
as well as with the total spin of the bath $\left|\vec{P}\right|$
. As discussed in the previous section of the Appendix, this result
is not correct for a local spin bath, because the Curie temperature
should not be affected by the longitudinally coupled bath. The correction
is provided by the correlations between the Ising spins and the bath
spins. On the other hand, this model can also have several Ising spins
coupled to each bath spin, and in this case the Curie temperature
can be affected. Similarly for the critical field one finds:
\begin{equation}
M_{z}\simeq\frac{Z_{B}A\left|\vec{P}\right|}{2\sqrt{B^{2}+A^{2}\left|\vec{P}\right|^{2}Z_{B}^{2}}}+\frac{B^{2}V_{0}M_{z}}{2\left[B^{2}+A^{2}\left|\vec{P}\right|^{2}Z_{B}^{2}\right]^{3/2}}+\mathcal{O}\left(M_{z}^{2}\right)
\end{equation}
Then $M_{z}=0$ is not a solution due to a remnant magnetization,
which saturates for large $A\gg B$.

\section{Effect of two-spin correlations}

In order to include two-point correlations, we include the Fock term
in the equation of motion. This couples Green's functions at different
lattice sites:
\begin{eqnarray*}
\omega G_{n,m}^{\alpha,\beta} & = & \langle\left[S_{n}^{\alpha},O_{m}^{\beta}\right]_{\pm}\rangle+i\sum_{\mu}\epsilon_{\mu\alpha\rho}\left(B_{\mu}+\sum_{i\neq n}V_{i,n}^{\mu}M_{i}^{\mu}-\sum_{l}A_{n,l}^{\mu}m_{l}^{\mu}\right)G_{n,m}^{\rho,\beta}\\
 &  & +i\sum_{\mu}\epsilon_{\mu\alpha\rho}M_{n}^{\rho}\left(\sum_{i\neq n}V_{i,n}^{\mu}G_{i,m}^{\mu,\beta}-\sum_{l}A_{n,l}^{\mu}Y_{l,m}^{\mu,\beta}\right)\\
\omega Y_{n,m}^{\alpha,\beta} & = & \langle\left[I_{n}^{\alpha},O_{m}^{\beta}\right]_{\pm}\rangle+i\sum_{\mu}\epsilon_{\mu\alpha\rho}\left(\Delta_{\mu}-\sum_{i}A_{i,n}^{\mu}\langle S_{i}^{\mu}\rangle\right)Y_{n,m}^{\rho,\beta}-i\sum_{\mu,i}\epsilon_{\mu\alpha\rho}A_{i,n}^{\mu}\langle I_{n}^{\rho}\rangle G_{i,m}^{\mu,\beta}
\end{eqnarray*}
Using a transformation to momentum space, while assuming spatially
homogeneous solutions, we find:
\begin{eqnarray}
\omega G_{k}^{\alpha,\beta} & = & \langle\left[S^{\alpha},O^{\beta}\right]_{\pm}\rangle+i\sum_{\mu}\epsilon_{\mu\alpha\rho}\omega_{s,\mu}G_{k}^{\rho,\beta}+i\sum_{\mu}\epsilon_{\mu\alpha\rho}M_{\rho}\left(V_{k}^{\mu}G_{k}^{\mu,\beta}-Z_{B}A^{\mu}Y_{k}^{\mu,\beta}\right)\\
\omega Y_{k}^{\alpha,\beta} & = & \langle\left[I^{\alpha},O^{\beta}\right]_{\pm}\rangle+i\sum_{\mu}\epsilon_{\mu\alpha\rho}\omega_{b,\mu}Y_{k}^{\rho,\beta}-i\sum_{\mu}\epsilon_{\mu\alpha\rho}m_{\rho}Z_{BS}A^{\mu}G_{k}^{\mu,\beta}\label{eq:BathEOM}
\end{eqnarray}
where $\omega_{s,\mu}=B_{\mu}+\sum_{i\neq n}V_{0}^{\mu}M_{\mu}-Z_{B}A^{\mu}m_{\mu}$
and $\omega_{b,\mu}=\Delta_{\mu}-Z_{BS}A^{\mu}M_{\mu}$.

For the case of the Ising model longitudinally coupled to a bath of
spins and $\Delta_{x}=0$, the poles of the Ising spins Green's functions
are at: 
\begin{eqnarray}
\omega_{k} & = & \sqrt{B^{2}+\left(M_{z}V_{0}-Z_{B}Am_{z}\right)^{2}-BM_{x}V_{k}}
\end{eqnarray}
The corresponding self-consistency equations are obtained from the
anti-commutator Green's functions:
\begin{eqnarray}
M_{z} & = & \frac{M_{z}V_{0}-Z_{B}Am_{z}}{2\frac{1}{N}\sum_{k}\omega_{k}\coth\left(\frac{\omega_{k}}{2T}\right)}\\
M_{x} & = & \frac{B}{2\frac{1}{N}\sum_{k}\omega_{k}\coth\left(\frac{\omega_{k}}{2T}\right)}
\end{eqnarray}
If the bath is static one can see that the bath occupation probabilities
are unchanged and that Ising spins and bath spins are not entangled.

If the bath has internal dynamics, now driven by $\Delta_{x}\neq0$,
one finds that the poles are not simple Ising magnons anymore, but
combinations of the bath and system excitations:
\begin{equation}
\omega_{\pm}=\sqrt{\frac{\Omega^{2}+\omega_{k}^{2}}{2}\pm\sqrt{\kappa+\left(\frac{\Omega^{2}-\omega_{k}^{2}}{2}\right)^{2}}}
\end{equation}
where $\kappa=Z_{B}Z_{BS}A^{2}B_{x}\Delta_{x}m_{x}M_{x}$, $\omega_{k}=\sqrt{B_{x}^{2}+\left(B_{z}+V_{0}M_{z}-Z_{B}Am_{z}\right)^{2}-B_{x}V_{k}M_{x}}$
and $\Omega=\sqrt{\Delta_{x}^{2}+\left(\Delta_{z}-Z_{BS}AM_{z}\right)}$.
Analogously one can calculate the self-consistency equations for the
magnetization, and the correlation functions:
\begin{eqnarray}
M_{z} & = & \frac{\omega_{z}}{2\xi},\ M_{x}=\frac{B_{x}}{2\xi}\\
\xi & \equiv & \frac{1}{N}\sum_{k}\frac{\omega_{+}\omega_{-}\left(\omega_{+}^{2}-\omega_{-}^{2}\right)}{\left(1-\lambda_{k}\right)\left[\left(\Omega^{2}-\omega_{-}^{2}\right)\omega_{+}\tanh\left(\frac{\omega_{-}}{2T}\right)-\left(\Omega^{2}-\omega_{+}^{2}\right)\omega_{-}\tanh\left(\frac{\omega_{+}}{2T}\right)\right]}\nonumber 
\end{eqnarray}
where
\begin{align*}
\lambda_{k} & =\frac{Z_{B}A\Delta_{x}M_{x}\left[\omega_{-}\tanh\left(\frac{\omega_{+}}{2T}\right)-\omega_{+}\tanh\left(\frac{\omega_{-}}{2T}\right)\right]}{\left(\Omega^{2}-\omega_{+}^{2}\right)\omega_{-}\tanh\left(\frac{\omega_{+}}{2T}\right)-\left(\Omega^{2}-\omega_{-}^{2}\right)\omega_{+}\tanh\left(\frac{\omega_{-}}{2T}\right)}\\
 & \times\frac{Z_{BS}AB_{x}m_{x}\left[\omega_{-}\tanh\left(\frac{\omega_{+}}{2T}\right)-\omega_{+}\tanh\left(\frac{\omega_{-}}{2T}\right)\right]}{\left(\omega_{k}^{2}-\omega_{+}^{2}\right)\omega_{-}\tanh\left(\frac{\omega_{+}}{2T}\right)-\left(\omega_{k}^{2}-\omega_{-}^{2}\right)\omega_{+}\tanh\left(\frac{\omega_{-}}{2T}\right)}
\end{align*}
Therefore the self-consistency equations can be written in an analogous
form to the case $\vec{\Delta}=0$, by just redefining the function
$\xi$, which contains a mixing term proportional to $\lambda_{k}\propto Z_{BS}Z_{B}A^{2}B_{x}\Delta_{x}M_{x}m_{x}$.
Importantly, in this case the bath-bath correlators are also non-vanishing,
as they get entangled through the Ising spins.

\section{Correlated parts}

The equations of motion for the correlated parts are obtained analogously,
by calculating the Heisenberg equations of motion:

\begin{eqnarray}
\partial_{t}S_{p}^{\rho}S_{n}^{\alpha} & = & \sum_{\mu,i}B_{\mu}\left(\epsilon_{\mu\rho\theta}S_{p}^{\theta}S_{n}^{\alpha}+\epsilon_{\mu\alpha\theta}S_{p}^{\rho}S_{n}^{\theta}\right)\nonumber \\
 &  & +\sum_{\mu}\sum_{i,j>i}V_{i,j}^{\mu}\left[\epsilon_{\mu\rho\theta}\left(\delta_{j,p}S_{i}^{\mu}S_{p}^{\theta}+\delta_{i,p}S_{p}^{\theta}S_{j}^{\mu}\right)S_{n}^{\alpha}+\epsilon_{\mu\alpha\theta}S_{p}^{\rho}\left(\delta_{j,n}S_{i}^{\mu}S_{n}^{\theta}+\delta_{i,n}S_{n}^{\theta}S_{j}^{\mu}\right)\right]\nonumber \\
 &  & -\sum_{\mu}\sum_{i,l}A_{i,l}^{\mu}I_{l}^{\mu}\left(\delta_{p,i}\epsilon_{\mu\rho\theta}S_{p}^{\theta}S_{n}^{\alpha}+\delta_{n,i}\epsilon_{\mu\alpha\theta}S_{p}^{\rho}S_{n}^{\theta}\right)
\end{eqnarray}
\begin{eqnarray}
\partial_{t}I_{p}^{\rho}S_{n}^{\alpha} & = & \sum_{\mu,\theta}\epsilon_{\mu\alpha\theta}B_{\mu}I_{p}^{\rho}S_{n}^{\theta}+\epsilon_{\mu\alpha\theta}\sum_{\mu,\theta}\sum_{i,j>i}V_{i,j}^{\mu}I_{p}^{\rho}\left(\delta_{j,n}S_{i}^{\mu}S_{n}^{\theta}+\delta_{i,n}S_{n}^{\theta}S_{j}^{\mu}\right)+\sum_{\mu,\theta}\epsilon_{\mu\rho\theta}\Delta_{\mu}I_{p}^{\theta}S_{n}^{\alpha}\nonumber \\
 &  & -\sum_{\mu}\sum_{i,l}A_{i,l}^{\mu}\left(\delta_{i,n}\epsilon_{\mu\alpha\theta}I_{l}^{\mu}I_{p}^{\rho}S_{n}^{\theta}+\delta_{l,p}\epsilon_{\mu\rho\theta}I_{p}^{\theta}S_{n}^{\alpha}S_{i}^{\mu}\right)\\
\partial_{t}I_{p}^{\rho}I_{n}^{\alpha} & = & \sum_{\mu}\left(\Delta_{\mu}-\sum_{i}S_{i}^{\mu}\right)\left(\epsilon_{\mu\rho\theta}A_{i,p}^{\mu}I_{p}^{\theta}I_{n}^{\alpha}+\epsilon_{\mu\alpha\theta}A_{i,n}^{\mu}I_{p}^{\rho}I_{n}^{\theta}\right)
\end{eqnarray}
The correlated part of the Green's functions are obtained subtracting
the uncorrelated contributions. For example for the case of $\mathcal{G}_{pn,m}^{\rho\alpha,\beta}$
one has $i\partial_{t}\mathcal{G}_{pn,m}^{\rho\alpha,\beta}=i\partial_{t}G_{pn,m}^{\rho\alpha,\beta}-i\langle S_{p}^{\rho}\rangle\partial_{t}G_{n,m}^{\alpha,\beta}-i\langle S_{n}^{\alpha}\rangle\partial_{t}G_{p,m}^{\rho,\beta}$.
Finally, separating into uncorrelated and correlated parts, and neglecting
higher order correlators, one finds:
\begin{eqnarray*}
\omega\mathcal{G}_{pn,m}^{\rho\alpha,\beta} & \simeq & \langle\left[S_{p}^{\rho}S_{n}^{\alpha},O_{m}^{\beta}\right]_{\pm}\rangle-\langle S_{p}^{\rho}\rangle\langle\left[S_{n}^{\alpha},O_{m}^{\beta}\right]_{\pm}\rangle-\langle S_{n}^{\alpha}\rangle\langle\left[S_{p}^{\rho},O_{m}^{\beta}\right]_{\pm}\rangle\\
 &  & +i\sum_{\mu,\theta}\epsilon_{\mu\rho\theta}B_{\mu}\mathcal{G}_{pn,m}^{\theta\alpha,\beta}+i\sum_{\mu,\theta}\epsilon_{\mu\alpha\theta}B_{\mu}\mathcal{G}_{pn,m}^{\rho\theta,\beta}\\
 &  & +i\sum_{\mu}\epsilon_{\mu\rho\theta}\sum_{i\neq p,n}V_{i,p}^{\mu}\left(\langle S_{i}^{\mu}S_{n}^{\alpha}\rangle_{c}G_{p,m}^{\theta,\beta}+\langle S_{n}^{\alpha}S_{p}^{\theta}\rangle_{c}G_{i,m}^{\mu,\beta}+\langle S_{i}^{\mu}\rangle\mathcal{G}_{pn,m}^{\theta\alpha,\beta}+\langle S_{p}^{\theta}\rangle\mathcal{G}_{in,m}^{\mu\alpha,\beta}\right)\\
 &  & +i\sum_{\mu}\epsilon_{\mu\alpha\theta}\sum_{i\neq p,n}V_{i,n}^{\mu}\left(\langle S_{i}^{\mu}S_{p}^{\rho}\rangle_{c}G_{n,m}^{\theta,\beta}+\langle S_{p}^{\rho}S_{n}^{\theta}\rangle_{c}G_{i,m}^{\mu,\beta}+\langle S_{i}^{\mu}\rangle\mathcal{G}_{pn,m}^{\rho\theta,\beta}+\langle S_{n}^{\theta}\rangle\mathcal{G}_{ip,m}^{\mu\rho,\beta}\right)\\
 &  & -i\sum_{\mu}\epsilon_{\mu\rho\theta}\sum_{l}A_{p,l}^{\mu}\left(\langle S_{n}^{\alpha}I_{l}^{\mu}\rangle_{c}G_{p,m}^{\theta,\beta}+\langle S_{n}^{\alpha}S_{p}^{\theta}\rangle_{c}Y_{l,m}^{\mu,\beta}+\langle I_{l}^{\mu}\rangle\mathcal{G}_{pn,m}^{\theta\alpha,\beta}+\langle S_{p}^{\theta}\rangle\mathcal{Y}_{ln,m}^{\mu\alpha,\beta}\right)\\
 &  & -i\sum_{\mu}\epsilon_{\mu\alpha\theta}\sum_{l}A_{n,l}^{\mu}\left(\langle S_{p}^{\rho}I_{l}^{\mu}\rangle_{c}G_{n,m}^{\theta,\beta}+\langle S_{p}^{\rho}S_{n}^{\theta}\rangle_{c}Y_{l,m}^{\mu,\beta}+\langle I_{l}^{\mu}\rangle\mathcal{G}_{pn,m}^{\rho\theta,\beta}+\langle S_{n}^{\theta}\rangle\mathcal{Y}_{lp,m}^{\mu\rho,\beta}\right)\\
 &  & +\frac{i}{4}\sum_{\theta}\epsilon_{\alpha\rho\theta}\left(V_{n,p}^{\alpha}G_{p,m}^{\theta,\beta}-V_{p,n}^{\rho}G_{n,m}^{\theta,\beta}\right)\\
 &  & -i\sum_{\mu,\theta}\epsilon_{\rho\mu\theta}V_{n,p}^{\theta}\left(\langle S_{n}^{\alpha}\rangle\langle S_{p}^{\mu}\rangle G_{n,m}^{\theta,\beta}+\langle S_{n}^{\alpha}\rangle\langle S_{n}^{\theta}\rangle G_{p,m}^{\mu,\beta}+\langle S_{n}^{\theta}\rangle\langle S_{p}^{\mu}\rangle G_{n,m}^{\alpha,\beta}\right)\\
 &  & +i\sum_{\mu,\theta}\epsilon_{\alpha\mu\theta}V_{p,n}^{\mu}\left(\langle S_{p}^{\rho}\rangle\langle S_{p}^{\mu}\rangle G_{n,m}^{\theta,\beta}+\langle S_{p}^{\rho}\rangle\langle S_{n}^{\theta}\rangle G_{p,m}^{\mu,\beta}+\langle S_{n}^{\theta}\rangle\langle S_{p}^{\mu}\rangle G_{p,m}^{\rho,\beta}\right)
\end{eqnarray*}
Similarly for the other Green's functions:
\begin{eqnarray*}
\omega\mathcal{Y}_{pn,m}^{\rho\alpha,\beta} & \simeq & \langle\left[I_{p}^{\rho}S_{n}^{\alpha},O_{m}^{\beta}\right]_{\pm}\rangle-\langle I_{p}^{\rho}\rangle\langle\left[S_{n}^{\alpha},O_{m}^{\beta}\right]_{\pm}\rangle-\langle S_{n}^{\alpha}\rangle\langle\left[I_{p}^{\rho},O_{m}^{\beta}\right]_{\pm}\rangle\\
 &  & +i\sum_{\mu,\theta}\left(\epsilon_{\mu\alpha\theta}B_{\mu}\mathcal{Y}_{pn,m}^{\rho\theta,\beta}+\epsilon_{\mu\rho\theta}\Delta_{\mu}\mathcal{Y}_{pn,m}^{\theta\alpha,\beta}\right)\\
 &  & +i\sum_{\mu,\theta}\epsilon_{\mu\alpha\theta}\sum_{i\neq n}V_{n,i}^{\mu}\left(\langle I_{p}^{\rho}S_{i}^{\mu}\rangle_{c}G_{n,m}^{\theta,\beta}+\langle I_{p}^{\rho}S_{n}^{\theta}\rangle_{c}G_{i,m}^{\mu,\beta}+\langle S_{n}^{\theta}\rangle\mathcal{Y}_{pi,m}^{\rho\mu,\beta}+\langle S_{i}^{\mu}\rangle\mathcal{Y}_{pn,m}^{\rho\theta,\beta}\right)\\
 &  & -i\sum_{\mu,\theta}\epsilon_{\mu\alpha\theta}\sum_{l\neq p}A_{n,l}^{\mu}\left(\langle I_{p}^{\rho}I_{l}^{\mu}\rangle_{c}G_{n,m}^{\theta,\beta}+\langle I_{p}^{\rho}S_{n}^{\theta}\rangle_{c}Y_{l,m}^{\mu,\beta}+\langle I_{l}^{\mu}\rangle\mathcal{Y}_{pn,m}^{\rho\theta,\beta}+\langle S_{n}^{\theta}\rangle\mathcal{W}_{lp,m}^{\mu\rho,\beta}\right)\\
 &  & -i\sum_{\mu,\theta}\epsilon_{\mu\rho\theta}\sum_{i\neq n}A_{i,p}^{\mu}\left(\langle I_{p}^{\theta}S_{i}^{\mu}\rangle G_{n,m}^{\alpha,\beta}+\langle I_{p}^{\theta}S_{n}^{\alpha}\rangle_{c}G_{i,m}^{\mu,\beta}+\langle S_{n}^{\alpha}S_{i}^{\mu}\rangle_{c}Y_{p,m}^{\theta,\beta}+\langle I_{p}^{\theta}\rangle\mathcal{G}_{in,m}^{\mu\alpha,\beta}+\langle S_{i}^{\mu}\rangle\mathcal{Y}_{pn,m}^{\theta\alpha,\beta}\right)\\
 &  & -i\sum_{\mu,\theta}\epsilon_{\mu\alpha\theta}A_{n,p}^{\mu}\left[\left(\langle I_{p}^{\mu}I_{p}^{\rho}\rangle-\langle I_{p}^{\rho}\rangle\langle I_{p}^{\mu}\rangle\right)G_{n,m}^{\theta,\beta}+\langle S_{n}^{\theta}\rangle\left(W_{pp,m}^{\mu\rho,\beta}-\langle I_{p}^{\rho}\rangle Y_{p,m}^{\mu,\beta}\right)\right]\\
 &  & -i\sum_{\mu,\theta}\epsilon_{\mu\rho\theta}A_{n,p}^{\mu}\left(\langle I_{p}^{\theta}\rangle G_{nn,m}^{\alpha\mu,\beta}-\langle S_{n}^{\alpha}\rangle\langle I_{p}^{\rho}\rangle G_{n,m}^{\mu,\beta}+\langle S_{n}^{\alpha}S_{n}^{\mu}\rangle Y_{p,m}^{\theta,\beta}-\langle S_{n}^{\alpha}\rangle\langle S_{n}^{\mu}\rangle Y_{p,m}^{\rho,\beta}\right)
\end{eqnarray*}
The two-bath spins correlation function is also needed:
\begin{eqnarray*}
\omega\mathcal{W}_{pn,m}^{\rho\alpha,\beta} & = & \langle\left[I_{p}^{\rho}I_{n}^{\alpha},O_{m}^{\beta}\right]_{\pm}\rangle-\langle I_{p}^{\rho}\rangle\langle\left[I_{n}^{\alpha},O_{m}^{\beta}\right]_{\pm}\rangle-\langle I_{n}^{\alpha}\rangle\langle\left[I_{p}^{\rho},O_{m}^{\beta}\right]_{\pm}\rangle\\
 &  & +i\sum_{\mu}\epsilon_{\mu\rho\theta}\left[\Delta_{\mu}\mathcal{W}_{pn,m}^{\theta\alpha,\beta}-\sum_{i}A_{i,p}^{\mu}\left(\langle I_{n}^{\alpha}I_{p}^{\theta}\rangle_{c}G_{i,m}^{\mu,\beta}+\langle I_{n}^{\alpha}S_{i}^{\mu}\rangle_{c}Y_{p,m}^{\theta,\beta}+\langle I_{p}^{\theta}\rangle\mathcal{Y}_{ni,m}^{\alpha\mu,\beta}+\langle S_{i}^{\mu}\rangle\mathcal{W}_{pn,m}^{\theta\alpha,\beta}\right)\right]\\
 &  & +i\sum_{\mu}\epsilon_{\mu\alpha\theta}\left[\Delta_{\mu}\mathcal{W}_{pn,m}^{\rho\theta,\beta}-\sum_{i}A_{i,n}^{\mu}\left(\langle I_{p}^{\rho}I_{n}^{\theta}\rangle_{c}G_{i,m}^{\mu,\beta}+\langle I_{p}^{\rho}S_{i}^{\mu}\rangle_{c}Y_{n,m}^{\theta,\beta}+\langle I_{n}^{\theta}\rangle\mathcal{Y}_{pi,m}^{\rho\mu,\beta}+\langle S_{i}^{\mu}\rangle\mathcal{W}_{pn,m}^{\rho\theta,\beta}\right)\right]
\end{eqnarray*}
These set of equations of motion provides very complex expressions
for the self-energy, and their role will be analyzed in future publications,
however to estimate their contribution, we consider the specific case
of the correction to the longitudinal magnetization. Furthermore,
we rewrite these equations of motion in terms of Majorana fermions
to simplify the self-consistency equations:
\begin{eqnarray}
S_{n}^{\alpha} & = & -\frac{i}{2}\epsilon_{\alpha\theta_{1}\theta_{2}}\eta_{n}^{\theta_{1}}\eta_{n}^{\theta_{2}}\label{eq:Majorana-Algebra}\\
I_{n}^{\alpha} & = & -\frac{i}{2}\epsilon_{\alpha\theta_{1}\theta_{2}}\gamma_{n}^{\theta_{1}}\gamma_{n}^{\theta_{2}}
\end{eqnarray}
$\eta_{n}^{\alpha}$ being a Majorana fermion at site $n$ that fulfills
the usual anti-commutation relation for fermions, and in addition
$\eta^{2}=1/2$. Similarly the $\gamma_{n}^{\alpha}$ refer to the
Majorana fermions for the bath spins. The Green's functions required
for the calculation of the magnetization $M_{n}^{\alpha}=-\frac{i}{2}\epsilon_{\alpha\theta_{1}\theta_{2}}\langle\eta_{n}^{\theta_{1}}\eta_{n}^{\theta_{2}}\rangle$
are:
\begin{eqnarray}
G_{n,m}^{\alpha,\beta}\left(t,t^{\prime}\right) & = & -i\langle\eta_{n}^{\alpha}\left(t\right);\eta_{m}^{\beta}\left(t^{\prime}\right)\rangle\\
P_{n,m}^{\alpha,\beta}\left(t,t^{\prime}\right) & = & -i\langle\gamma_{n}^{\alpha}\left(t\right);\gamma_{m}^{\beta}\left(t^{\prime}\right)\rangle
\end{eqnarray}
The longitudinal magnetization is obtained from the diagonal two-point
function, with equation of motion:
\begin{eqnarray}
\omega G_{n,n}^{\alpha,\beta} & = & \delta_{\alpha,\beta}+i\epsilon_{\mu\alpha\theta}h_{s}^{\mu}\left(n\right)G_{n,n}^{\theta,\beta}\label{eq:EOM-G}\\
 &  & +\frac{1}{2}\epsilon_{\mu\alpha\theta}\epsilon_{\nu\theta_{1}\theta_{2}}\sum_{i\neq n}V_{n,i}^{\mu,\nu}\mathcal{G}_{iin,n}^{\theta_{1}\theta_{2}\theta,\beta}\nonumber \\
 &  & -\frac{1}{2}\epsilon_{\mu\alpha\theta}\epsilon_{\nu\theta_{1}\theta_{2}}\sum_{l}A_{n,l}^{\mu,\nu}\mathcal{Y}_{lln,n}^{\theta_{1}\theta_{2}\theta,\beta}\nonumber 
\end{eqnarray}
which also requires the calculation of the bath spins Green's function:
\begin{eqnarray}
\omega P_{l,l}^{\alpha,\beta} & = & \delta_{\alpha,\beta}+i\epsilon_{\mu\alpha\theta}h_{b}^{\mu}\left(l\right)P_{l,l}^{\theta,\beta}\label{eq:EOM-P}\\
 &  & -\frac{1}{2}\epsilon_{\mu\theta_{1}\theta_{2}}\epsilon_{\nu\alpha\theta}\sum_{i}A_{i,l}^{\mu,\nu}\mathcal{R}_{iil,l}^{\theta_{1}\theta_{2}\theta,\beta}\nonumber 
\end{eqnarray}
where we have defined $h_{s}^{\mu}\left(n\right)=B_{\mu}+\sum_{i\neq n}V_{n,i}^{\mu,\nu}M_{i}^{\nu}-\sum_{l}A_{n,l}^{\mu,\nu}m_{l}^{\nu}$,
$h_{b}^{\mu}\left(l\right)=\mathcal{B}_{\mu}-\sum_{i}A_{i,l}^{\nu,\mu}M_{i}^{\nu}$,
and $\mathcal{Y}_{lln,n}^{\theta_{1}\theta_{2}\theta,\beta}$ and
$\mathcal{R}_{iil,l}^{\theta_{1}\theta_{2}\theta,\beta}$ are the
correlated parts of the mixed-Green's functions $Y_{lln,n}^{\theta_{1}\theta_{2}\theta,\beta}=-i\langle\gamma_{l}^{\theta_{3}}\gamma_{l}^{\theta_{4}}\eta_{n}^{\theta_{1}};\eta_{n}^{\beta}\rangle$
and $R_{iil,l}^{\theta_{1}\theta_{2}\theta,\beta}=-i\langle\eta_{i}^{\theta_{1}}\eta_{i}^{\theta_{2}}\gamma_{l}^{\theta};\gamma_{l}^{\beta}\rangle$,
respectively.

To lowest order one recovers the MF expressions presented in the first
sections of the main text. In this case, as we are calculating the
diagonal Green's function $G_{n,n}^{\alpha,\beta}$, there is no contribution
from a Fock term to lowest order, and the contribution from magnons
comes in from the correlated parts. Their equations of motion are:
\begin{eqnarray*}
\omega\mathcal{G}_{ppn,n}^{\sigma_{1}\sigma_{2}\alpha,\beta} & = & -i\sum_{\mu}B_{\mu}\left(\epsilon_{\mu\theta\sigma_{1}}\mathcal{G}_{ppn,n}^{\theta\sigma_{2}\alpha,\beta}+\epsilon_{\mu\theta\sigma_{2}}\mathcal{G}_{ppn,n}^{\sigma_{1}\theta\alpha,\beta}+\epsilon_{\mu\theta\alpha}\mathcal{G}_{ppn,n}^{\sigma_{1}\sigma_{2}\theta,\beta}\right)\\
 &  & +\frac{1}{2}\sum_{\mu,\nu}\epsilon_{\nu\theta_{1}\theta_{2}}\sum_{k}A_{p,k}^{\mu,\nu}\epsilon_{\mu\theta\sigma_{1}}\left(\langle\gamma_{k}^{\theta_{1}}\gamma_{k}^{\theta_{2}}\rangle\mathcal{G}_{ppn,n}^{\theta\sigma_{2}\alpha,\beta}+\langle\eta_{p}^{\theta}\eta_{p}^{\sigma_{2}}\rangle\mathcal{Y}_{kkn,n}^{\theta_{1}\theta_{2}\alpha,\beta}\right)\\
 &  & +\frac{1}{2}\sum_{\mu,\nu}\epsilon_{\nu\theta_{1}\theta_{2}}\sum_{k}A_{p,k}^{\mu,\nu}\epsilon_{\mu\theta\sigma_{2}}\left(\langle\gamma_{k}^{\theta_{1}}\gamma_{k}^{\theta_{2}}\rangle\mathcal{G}_{ppn,n}^{\sigma_{1}\theta\alpha,\beta}+\langle\eta_{p}^{\sigma_{1}}\eta_{p}^{\theta}\rangle\mathcal{Y}_{kkn,n}^{\theta_{1}\theta_{2}\alpha,\beta}\right)\\
 &  & +\frac{1}{2}\sum_{\mu,\nu}\epsilon_{\nu\theta_{1}\theta_{2}}\sum_{k}A_{n,k}^{\mu,\nu}\epsilon_{\mu\theta\alpha}\left(\langle\gamma_{k}^{\theta_{1}}\gamma_{k}^{\theta_{2}}\eta_{p}^{\sigma_{1}}\eta_{p}^{\sigma_{2}}\rangle_{C}G_{n,n}^{\theta,\beta}+\langle\gamma_{k}^{\theta_{1}}\gamma_{k}^{\theta_{2}}\rangle\mathcal{G}_{ppn,n}^{\sigma_{1}\sigma_{2}\theta,\beta}\right)\\
 &  & -\frac{1}{2}\sum_{\mu,\nu}\epsilon_{\mu\theta_{1}\theta_{2}}\sum_{i\neq p,n}V_{i,p}^{\mu,\nu}\epsilon_{\nu\theta\sigma_{1}}\left(\langle\eta_{i}^{\theta_{1}}\eta_{i}^{\theta_{2}}\rangle\mathcal{G}_{ppn,n}^{\theta\sigma_{2}\alpha,\beta}+\langle\eta_{p}^{\theta}\eta_{p}^{\sigma_{2}}\rangle\mathcal{G}_{iin,n}^{\theta_{1}\theta_{2}\alpha,\beta}\right)\\
 &  & -\frac{1}{2}\sum_{\mu,\nu}\epsilon_{\mu\theta_{1}\theta_{2}}\sum_{i\neq p,n}V_{i,p}^{\mu,\nu}\epsilon_{\nu\theta\sigma_{2}}\left(\langle\eta_{i}^{\theta_{1}}\eta_{i}^{\theta_{2}}\rangle\mathcal{G}_{ppn,n}^{\sigma_{1}\theta\alpha,\beta}+\langle\eta_{p}^{\sigma_{1}}\eta_{p}^{\theta}\rangle\mathcal{G}_{iin,n}^{\theta_{1}\theta_{2}\alpha,\beta}\right)\\
 &  & -\frac{1}{2}\sum_{\mu,\nu}\epsilon_{\mu\theta_{1}\theta_{2}}\sum_{i\neq p,n}V_{i,n}^{\mu,\nu}\epsilon_{\nu\theta\alpha}\left(\langle\eta_{p}^{\sigma_{1}}\eta_{p}^{\sigma_{2}}\eta_{i}^{\theta_{1}}\eta_{i}^{\theta_{2}}\rangle_{C}G_{n,n}^{\theta,\beta}+\langle\eta_{i}^{\theta_{1}}\eta_{i}^{\theta_{2}}\rangle\mathcal{G}_{ppn,n}^{\sigma_{1}\sigma_{2}\theta,\beta}\right)\\
 &  & -\frac{1}{2}\sum_{\mu,\nu}\epsilon_{\mu\theta_{1}\theta_{2}}V_{n,p}^{\mu,\nu}\left(\epsilon_{\nu\theta\sigma_{1}}\langle\eta_{p}^{\theta}\eta_{p}^{\sigma_{2}}\rangle+\epsilon_{\nu\theta\sigma_{2}}\langle\eta_{p}^{\sigma_{1}}\eta_{p}^{\theta}\rangle\right)\left(G_{nnn,n}^{\theta_{1}\theta_{2}\alpha,\beta}-\langle\eta_{n}^{\theta_{1}}\eta_{n}^{\theta_{2}}\rangle G_{n,n}^{\alpha,\beta}\right)\\
 &  & -\frac{1}{2}\sum_{\mu,\nu}\epsilon_{\mu\theta_{1}\theta_{2}}\epsilon_{\nu\theta\alpha}V_{p,n}^{\mu,\nu}\left(\langle\eta_{p}^{\sigma_{1}}\eta_{p}^{\sigma_{2}}\eta_{p}^{\theta_{1}}\eta_{p}^{\theta_{2}}\rangle-\langle\eta_{p}^{\sigma_{1}}\eta_{p}^{\sigma_{2}}\rangle\langle\eta_{p}^{\theta_{1}}\eta_{p}^{\theta_{2}}\rangle\right)G_{n,n}^{\theta,\beta}
\end{eqnarray*}
\begin{eqnarray*}
\omega\mathcal{Y}_{lln,n}^{\sigma_{1}\sigma_{2}\alpha,\beta} & = & -i\sum_{\mu}B_{\mu}\epsilon_{\mu\theta\alpha}\mathcal{Y}_{lln,n}^{\sigma_{1}\sigma_{2}\theta,\beta}-i\sum_{\mu}\Delta_{\mu}\left(\epsilon_{\mu\theta\sigma_{1}}\mathcal{Y}_{lln,n}^{\theta\sigma_{2}\alpha,\beta}+\epsilon_{\mu\theta\sigma_{2}}\mathcal{Y}_{lln,n}^{\sigma_{1}\theta\alpha,\beta}\right)\\
 &  & -\frac{1}{2}\sum_{\mu,\nu}\epsilon_{\mu\theta_{1}\theta_{2}}\epsilon_{\nu\theta\alpha}\sum_{i\neq n}V_{i,n}^{\mu,\nu}\left(\langle\gamma_{l}^{\sigma_{1}}\gamma_{l}^{\sigma_{2}}\eta_{i}^{\theta_{1}}\eta_{i}^{\theta_{2}}\rangle^{C}G_{n,n}^{\theta,\beta}+\langle\eta_{i}^{\theta_{1}}\eta_{i}^{\theta_{2}}\rangle\mathcal{Y}_{lln,n}^{\sigma_{1}\sigma_{2}\theta,\beta}\right)\\
 &  & +\frac{1}{2}\sum_{\mu,\nu}\epsilon_{\nu\theta_{1}\theta_{2}}\epsilon_{\mu\theta\alpha}\sum_{r\neq l}A_{n,r}^{\mu,\nu}\left(\langle\gamma_{l}^{\sigma_{1}}\gamma_{l}^{\sigma_{2}}\gamma_{r}^{\theta_{1}}\gamma_{r}^{\theta_{2}}\rangle^{C}G_{n,n}^{\theta,\beta}+\langle\gamma_{r}^{\theta_{1}}\gamma_{r}^{\theta_{2}}\rangle\mathcal{Y}_{lln,n}^{\sigma_{1}\sigma_{2}\theta,\beta}\right)\\
 &  & +\frac{1}{2}\sum_{\mu,\nu}\epsilon_{\mu\theta_{1}\theta_{2}}\sum_{i\neq n}A_{i,l}^{\mu,\nu}\epsilon_{\nu\theta\sigma_{1}}\left(\langle\gamma_{l}^{\theta}\gamma_{l}^{\sigma_{2}}\rangle\mathcal{G}_{iin,n}^{\theta_{1}\theta_{2}\alpha,\beta}+\langle\eta_{i}^{\theta_{1}}\eta_{i}^{\theta_{2}}\rangle\mathcal{Y}_{lln,n}^{\theta\sigma_{2}\alpha,\beta}\right)\\
 &  & +\frac{1}{2}\sum_{\mu,\nu}\epsilon_{\mu\theta_{1}\theta_{2}}\sum_{i\neq n}A_{i,l}^{\mu,\nu}\epsilon_{\nu\theta\sigma_{2}}\left(\langle\gamma_{l}^{\sigma_{1}}\gamma_{l}^{\theta}\rangle\mathcal{G}_{iin,n}^{\theta_{1}\theta_{2}\alpha,\beta}+\langle\eta_{i}^{\theta_{1}}\eta_{i}^{\theta_{2}}\rangle\mathcal{Y}_{lln,n}^{\sigma_{1}\theta\alpha,\beta}\right)\\
 &  & +\frac{1}{2}\sum_{\mu,\nu}\epsilon_{\mu\theta_{1}\theta_{2}}A_{n,l}^{\mu,\nu}\left(\epsilon_{\nu\theta\sigma_{1}}\langle\gamma_{l}^{\theta}\gamma_{l}^{\sigma_{2}}\rangle+\epsilon_{\nu\theta\sigma_{2}}\langle\gamma_{l}^{\sigma_{1}}\gamma_{l}^{\theta}\rangle\right)G_{nnn,n}^{\theta_{1}\theta_{2}\alpha,\beta}\\
 &  & +\frac{1}{2}\sum_{\mu,\nu}\epsilon_{\nu\theta_{1}\theta_{2}}\epsilon_{\mu\theta\alpha}A_{n,l}^{\mu,\nu}\left(\langle\gamma_{l}^{\sigma_{1}}\gamma_{l}^{\sigma_{2}}\gamma_{l}^{\theta_{1}}\gamma_{l}^{\theta_{2}}\rangle-\langle\gamma_{l}^{\sigma_{1}}\gamma_{l}^{\sigma_{2}}\rangle\langle\gamma_{l}^{\theta_{1}}\gamma_{l}^{\theta_{2}}\rangle\right)G_{n,n}^{\theta,\beta}
\end{eqnarray*}
\begin{eqnarray*}
\omega\mathcal{R}_{ppl,l}^{\sigma_{1}\sigma_{2}\alpha,\beta} & = & -i\sum_{\mu}B_{\mu}\left(\epsilon_{\mu\theta\sigma_{1}}\mathcal{R}_{ppl,l}^{\theta\sigma_{2}\alpha,\beta}+\epsilon_{\mu\theta\sigma_{2}}\mathcal{R}_{ppl,l}^{\sigma_{1}\theta\alpha,\beta}\right)-i\sum_{\mu}\Delta_{\mu}\epsilon_{\mu\theta\alpha}\mathcal{R}_{ppl,l}^{\sigma_{1}\sigma_{2}\theta,\beta}\\
 &  & -\frac{1}{2}\sum_{\mu,\nu}\epsilon_{\mu\theta_{1}\theta_{2}}\sum_{i\neq p}V_{i,p}^{\mu,\nu}\epsilon_{\nu\theta\sigma_{1}}\left(\langle\eta_{i}^{\theta_{1}}\eta_{i}^{\theta_{2}}\rangle\mathcal{R}_{ppl,l}^{\theta\sigma_{2}\alpha,\beta}+\langle\eta_{p}^{\theta}\eta_{p}^{\sigma_{2}}\rangle\mathcal{R}_{iil,l}^{\theta_{1}\theta_{2}\alpha,\beta}\right)\\
 &  & -\frac{1}{2}\sum_{\mu,\nu}\epsilon_{\mu\theta_{1}\theta_{2}}\sum_{i\neq p}V_{i,p}^{\mu,\nu}\epsilon_{\nu\theta\sigma_{2}}\left(\langle\eta_{i}^{\theta_{1}}\eta_{i}^{\theta_{2}}\rangle\mathcal{R}_{ppl,l}^{\sigma_{1}\theta\alpha,\beta}+\langle\eta_{p}^{\sigma_{1}}\eta_{p}^{\theta}\rangle\mathcal{R}_{iil,l}^{\theta_{1}\theta_{2}\alpha,\beta}\right)\\
 &  & +\frac{1}{2}\sum_{\mu,\nu}\epsilon_{\mu\theta_{1}\theta_{2}}\sum_{i\neq p}A_{i,l}^{\mu,\nu}\epsilon_{\nu\theta\alpha}\left(\langle\eta_{p}^{\sigma_{1}}\eta_{p}^{\sigma_{2}}\eta_{i}^{\theta_{1}}\eta_{i}^{\theta_{2}}\rangle^{C}P_{l,l}^{\theta,\beta}+\langle\eta_{i}^{\theta_{1}}\eta_{i}^{\theta_{2}}\rangle\mathcal{R}_{ppl,l}^{\sigma_{1}\sigma_{2}\theta,\beta}\right)\\
 &  & +\frac{1}{2}\sum_{\mu,\nu}\epsilon_{\nu\theta_{1}\theta_{2}}\sum_{r\neq l}A_{p,r}^{\mu,\nu}\epsilon_{\mu\theta\sigma_{1}}\left(\langle\eta_{p}^{\theta}\eta_{p}^{\sigma_{2}}\rangle\mathcal{P}_{rrl,l}^{\theta_{1}\theta_{2}\alpha,\beta}+\langle\gamma_{r}^{\theta_{1}}\gamma_{r}^{\theta_{2}}\rangle\mathcal{R}_{ppl,l}^{\theta\sigma_{2}\alpha,\beta}\right)\\
 &  & +\frac{1}{2}\sum_{\mu,\nu}\epsilon_{\nu\theta_{1}\theta_{2}}\sum_{r\neq l}A_{p,r}^{\mu,\nu}\epsilon_{\mu\theta\sigma_{2}}\left(\langle\eta_{p}^{\sigma_{1}}\eta_{p}^{\theta}\rangle\mathcal{P}_{rrl,l}^{\theta_{1}\theta_{2}\alpha,\beta}+\langle\gamma_{r}^{\theta_{1}}\gamma_{r}^{\theta_{2}}\rangle\mathcal{R}_{ppl,l}^{\sigma_{1}\theta\alpha,\beta}\right)\\
 &  & +\frac{1}{2}\sum_{\mu,\nu}\epsilon_{\mu\theta_{1}\theta_{2}}A_{p,l}^{\mu,\nu}\epsilon_{\nu\theta\alpha}\left(\langle\eta_{p}^{\theta_{1}}\eta_{p}^{\theta_{2}}\eta_{p}^{\sigma_{1}}\eta_{p}^{\sigma_{2}}\rangle-\langle\eta_{p}^{\sigma_{1}}\eta_{p}^{\sigma_{2}}\rangle\langle\eta_{p}^{\theta_{1}}\eta_{p}^{\theta_{2}}\rangle\right)P_{l,l}^{\theta,\beta}\\
 &  & +\frac{1}{2}\sum_{\mu,\nu}\epsilon_{\nu\theta_{1}\theta_{2}}A_{p,l}^{\mu,\nu}\left(\epsilon_{\mu\theta\sigma_{1}}\langle\eta_{p}^{\theta}\eta_{p}^{\sigma_{2}}\rangle+\epsilon_{\mu\theta\sigma_{2}}\langle\eta_{p}^{\sigma_{1}}\eta_{p}^{\theta}\rangle\right)\left(P_{lll,l}^{\alpha\theta_{1}\theta_{2},\beta}-\langle\gamma_{l}^{\theta_{1}}\gamma_{l}^{\theta_{2}}\rangle P_{l,l}^{\alpha,\beta}\right)
\end{eqnarray*}
\begin{eqnarray*}
\omega\mathcal{P}_{rrl,l}^{\sigma_{1}\sigma_{2}\alpha,\beta} & = & -i\sum_{\mu}\Delta_{\mu}\left(\epsilon_{\mu\theta\sigma_{1}}\mathcal{P}_{rrl,l}^{\theta\sigma_{2}\alpha,\beta}+\epsilon_{\mu\theta\sigma_{2}}\mathcal{P}_{rrl,l}^{\sigma_{1}\theta\alpha,\beta}+\epsilon_{\mu\theta\alpha}\mathcal{P}_{rrl,l}^{\sigma_{1}\sigma_{2}\theta,\beta}\right)\\
 &  & +\frac{1}{2}\sum_{\mu,\nu}\epsilon_{\mu\theta_{1}\theta_{2}}\sum_{i}A_{i,r}^{\mu,\nu}\epsilon_{\nu\theta\sigma_{1}}\left(\langle\eta_{i}^{\theta_{1}}\eta_{i}^{\theta_{2}}\rangle\mathcal{P}_{rrl,l}^{\theta\sigma_{2}\alpha,\beta}+\langle\gamma_{r}^{\theta}\gamma_{r}^{\sigma_{2}}\rangle\mathcal{R}_{iil,l}^{\theta_{1}\theta_{2}\alpha,\beta}\right)\\
 &  & +\frac{1}{2}\sum_{\mu,\nu}\epsilon_{\mu\theta_{1}\theta_{2}}\sum_{i}A_{i,r}^{\mu,\nu}\epsilon_{\nu\theta\sigma_{2}}\left(\langle\eta_{i}^{\theta_{1}}\eta_{i}^{\theta_{2}}\rangle\mathcal{P}_{rrl,l}^{\sigma_{1}\theta\alpha,\beta}+\langle\gamma_{r}^{\sigma_{1}}\gamma_{r}^{\theta}\rangle\mathcal{R}_{iil,l}^{\theta_{1}\theta_{2}\alpha,\beta}\right)\\
 &  & +\frac{1}{2}\sum_{\mu,\nu}\epsilon_{\mu\theta_{1}\theta_{2}}\epsilon_{\nu\theta\alpha}\sum_{i}A_{i,l}^{\mu,\nu}\left(\langle\eta_{i}^{\theta_{1}}\eta_{i}^{\theta_{2}}\gamma_{r}^{\sigma_{1}}\gamma_{r}^{\sigma_{2}}\rangle^{C}P_{l,l}^{\theta,\beta}+\langle\eta_{i}^{\theta_{1}}\eta_{i}^{\theta_{2}}\rangle\mathcal{P}_{rrl,l}^{\sigma_{1}\sigma_{2}\theta,\beta}\right)
\end{eqnarray*}
with the addition of the exact equation: 
\begin{equation}
G_{nnn,n}^{xyz,\beta}\left(\omega\right)=\frac{i}{\omega}\left(\delta_{x,\beta}M_{x}+\delta_{y,\beta}M_{y}+\delta_{z,\beta}M_{z}\right)
\end{equation}
These equations are exactly solved for the case of the quantum Ising
model in the main text. Finally, we solve the self-consistency equations
numerically, until full convergence. This produces the phase diagram
shown in Fig.\ref{fig:Magnetization-Comparison}.

\end{widetext} 
\end{document}